\documentclass[10pt]{article}
\usepackage{pst-all}
\usepackage{letterspace}
\usepackage{psfrag}
\usepackage{latexsym}
\usepackage{amsmath}
\usepackage{amssymb}
\usepackage[dvips]{graphicx}
\usepackage{caption2}
\usepackage[noend]{algorithmic}
\usepackage[plain]{algorithm}

\usepackage{color}
\usepackage[nohead, top=3cm, bottom=1cm, left=2.9cm,
right=3.1cm]{geometry}

\usepackage{letterspace}
\usepackage{graphicx}
\usepackage{psfrag}

\def\yskip{\penalty-50\vskip3pt plus3pt minus2pt}
\def\y{\yskip}
\def\yy{\yskip\yskip}
\def\yyy{\yskip\yskip\yskip}
\def\qed{\hskip 3pt\vrule height 6pt width 3pt depth 0pt}
\def\q5uad{\quad\quad\quad\quad\quad}

\def\s{\ }

\usepackage{fancyheadings}
\title{\bf The 1-fixed-endpoint Path Cover Problem is Polynomial on Interval Graphs \vspace{0.4cm}}

\author{\large Katerina Asdre \s and \s Stavros D. Nikolopoulos}

\date{}

\begin{document}

\maketitle

\vspace{-0.6cm}

\centerline{\it Department of Computer Science, University of
Ioannina}

\centerline{\it P.O.Box 1186, GR-45110 \s Ioannina, Greece}

\centerline{\tt \{katerina, stavros\}@cs.uoi.gr}

\vskip 0.3in

\begin{center}
\noindent
\parbox{5.5in}
{{\bf Abstract:} \s We consider a variant of the path cover
problem, namely, the $k$-fixed-endpoint path cover problem, or kPC
for short, on interval graphs. Given a graph $G$ and a subset
$\mathcal{T}$ of $k$ vertices of $V(G)$, a $k$-fixed-endpoint path
cover of $G$ with respect to $\mathcal{T}$ is a set of
vertex-disjoint paths $\mathcal{P}$ that covers the vertices of
$G$ such that the $k$ vertices of $\mathcal{T}$ are all endpoints
of the paths in $\mathcal{P}$. The kPC problem is to find a
$k$-fixed-endpoint path cover of $G$ of minimum cardinality; note
that, if $\mathcal{T}$ is empty the stated problem coincides with
the classical path cover problem. In this paper, we study the
1-fixed-endpoint path cover problem on interval graphs, or 1PC for
short, generalizing the 1HP problem which has been proved to be
NP-complete even for small classes of graphs. Motivated by a work
of Damaschke \cite{Damaschke}, where he left both 1HP and 2HP
problems open for the class of interval graphs, we show that the
1PC problem can be solved in polynomial time on the class of
interval graphs. The proposed algorithm is simple, runs in
$O(n^2)$ time, requires linear space, and also enables us to solve
the 1HP problem on interval graphs within the same time and space
complexity.

\bigskip
\noindent {\bf Keywords:} \s perfect graphs, interval graphs, path
cover, fixed-endpoint path cover, linear-time algorithms.}
\end{center}

\vskip 0.3in 
\section{Introduction}
{\bf Framework--Motivation.} A well studied problem with numerous
practical applications in graph theory is to find a minimum number
of vertex-disjoint paths of a graph $G$ that cover the vertices of
$G$. This problem, also known as the path cover problem (PC),
finds application in the fields of database design, networks, code
optimization among many others (see \cite{AdPe90, AR90, LiOlPru95,
SSSR93}); it is well known that the path cover problem and many of
its variants are NP-complete in general graphs \cite{GaJo79}. A
graph that admits a path cover of size one is referred to as
Hamiltonian. Thus, the path cover problem is at least as hard as
the Hamiltonian path problem (HP), that is, the problem of
deciding whether a graph is Hamiltonian. The path cover problem is
known to be NP-complete even when the input is restricted to
several interesting special classes of graphs; for example, it is
NP-complete on planar graphs \cite{GaJoTar}, bipartite graphs
\cite{Gol}, chordal graphs \cite{Gol}, chordal bipartite graphs
\cite{Muller} and strongly chordal graphs \cite{Muller}. Bertossi
and Bonuccelli \cite{BertBonucc} proved that the Hamiltonian
Circuit problem is NP-complete on several interesting classes of
intersection graphs.

\y Several variants of the HP problem are also of great interest,
among which is the problem of deciding whether a graph admits a
Hamiltonian path between two points (2HP). The 2HP problem is the
same as the HP problem except that in 2HP two vertices of the
input graph~$G$ are specified, say, $u$ and $v$,  and we are asked
whether $G$ contains a Hamiltonian path beginning with $u$ and
ending with $v$. Similarly, the 1HP problem is to determine
whether a graph~$G$ admits a Hamiltonian path starting from a
specific vertex $u$ of $G$, and to find one if such a path does
exist. Both 1HP and 2HP problems are also NP-complete in general
graphs \cite{GaJo79}. In \cite{Damaschke}, Damaschke provided a
foundation for obtaining polynomial-time algorithms for several
problems concerning paths in interval graphs, such as finding
Hamiltonian paths and circuits, and partitions into paths. In the
same paper, he stated that the complexity status of both 1HP and
2HP problems on interval graphs remains an open question.
Motivated by the above issues we state a variant of the path cover
problem, namely, the 1-fixed-endpoint path cover problem (1PC),
which generalizes the 1HP problem.

\yy\noindent Problem 1PC: Given a graph $G$ and a vertex $u \in
V(G)$, a {\it 1-fixed-endpoint path cover} of the graph~$G$ with
respect to $u$ is a path cover of $G$ such that the vertex $u$ is
an endpoint of a path in the path cover; a {\it minimum
1-fixed-endpoint path cover} of $G$ with respect to $u$ is a
1-fixed-endpoint path cover of $G$ with minimum cardinality; the
{\it 1-fixed-endpoint path cover problem} (1PC) is to find a
minimum 1-fixed-endpoint path cover of the graph~$G$.

\begin{figure}[t]
\yy \hrule \y\y\y
  \centering
  \includegraphics[scale=0.7]{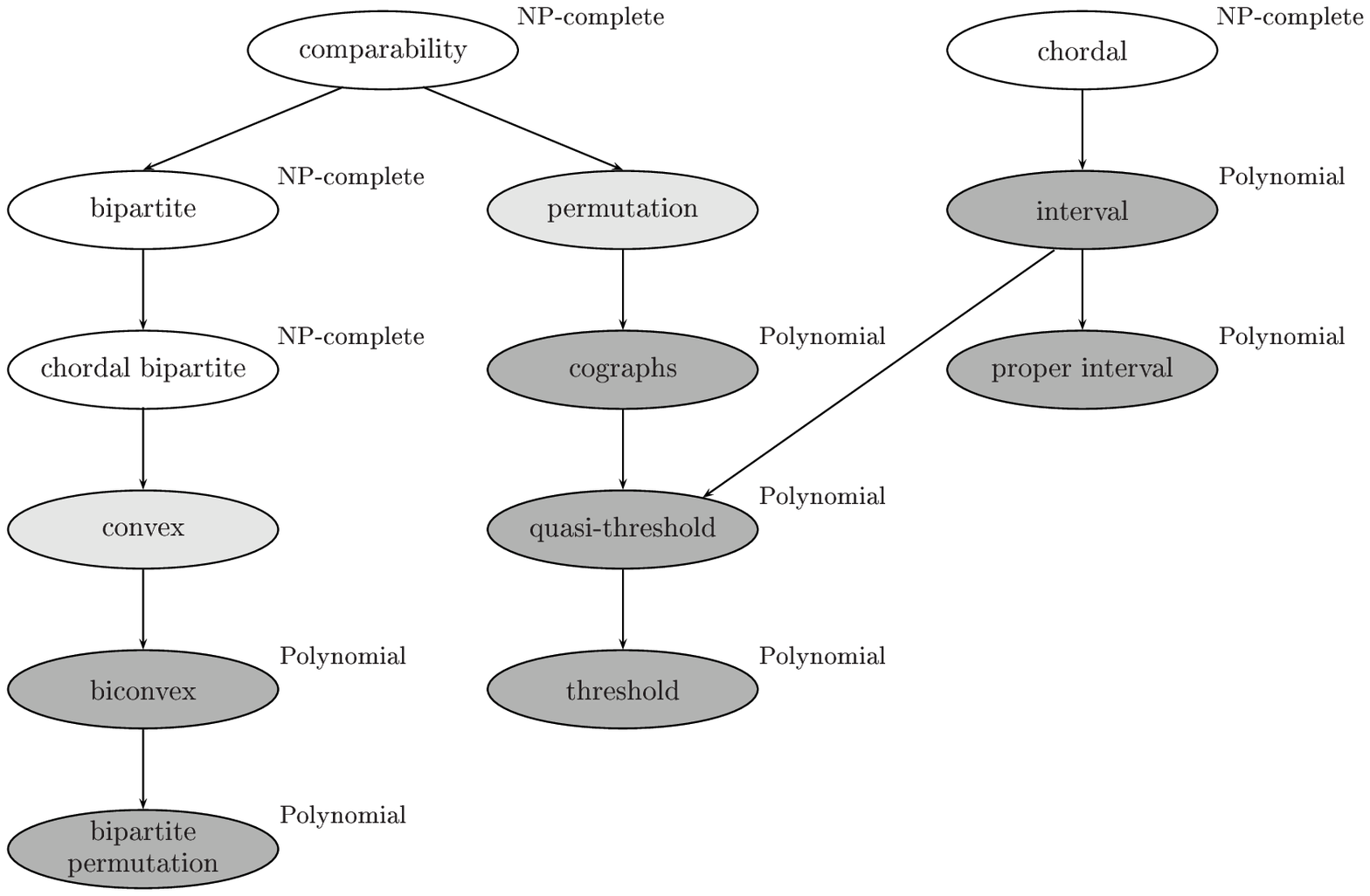}
  \centering
  \caption{\small{The complexity status (NP-complete, unknown, polynomial) of the 1HP problem for some graph subclasses of comparability and chordal graphs. $A \rightarrow B$ indicates that
  class $A$ contains class $B$.}}
  \label{class}
 \yy \hrule \y\y
\end{figure}

\yy\noindent {\bf Contribution.} In this paper, we study the
complexity status of the 1-fixed-endpoint path cover problem (1PC)
on the class of interval graphs \cite{BraLeSpi, Gol}, and show
that this problem can be solved in polynomial time. The proposed
algorithm runs in $O(n^2)$ time on an interval graph~$G$ on $n$
vertices and $m$ edges and requires linear space. The proposed
algorithm for the 1PC problem can also be used to solve the 1HP
problem on interval graphs within the same time and space
complexity. Using our algorithm for the 1PC problem and a simple
reduction described by M\"{u}ller in \cite{Muller}, we solve the
HP problem on a $X$-convex graph $G=(X,Y,E)$ with $|Y|-|X|=1$,
which was left open in \cite{UeharaUno}. We also show that the 1HP
problem on a convex graph $G$ is solvable in time quadratic in the
number of its vertices. Figure~\ref{class} shows a diagram of
class inclusions for a number of graph classes, subclasses of
comparability and chordal graphs, and the current complexity
status of the 1HP problem on these classes; for definitions of the
classes shown, see \cite{BraLeSpi, Gol}.

\yy\noindent {\bf Related Work.} Interval graphs form an important
class of perfect graphs \cite{Gol} and many problems that are
NP-complete on arbitrary graphs are shown to admit polynomial time
algorithms on this class \cite{AR90, Gol, Keil}. Both Hamiltonian
Circuit (HC) and Hamiltonian Path (HP) problems are polynomially
solvable for the class of interval and proper interval graphs.
Keil introduced a linear-time algorithm for the HC problem on
interval graphs \cite{Keil} and Arikati and Rangan \cite{AR90}
presented a linear-time algorithm for the minimum path cover
problem on interval graphs. Bertossi \cite{Bertossi} proved that a
proper interval graph has a Hamiltonian path if and only if it is
connected. He also gave an $O(n\log n)$ algorithm for finding a
Hamiltonian circuit in a proper interval graph. Recently, Asdre
and Nikolopoulos proposed a linear-time algorithm for the
k-fixed-endpoint path cover problem (kPC) on cographs and on
proper interval graphs \cite{AsdNik, AsdNik2}. Furthermore, Lin
et~al. \cite{LiOlPru95} proposed an optimal algorithm for the path
cover problem on cographs while Nakano et~al. \cite{NaOlZo03}
proposed an optimal parallel algorithm which finds and reports all
the paths in a minimum path cover of a cograph in $O(\log n)$ time
using $O(n/\log n)$ processors on a PRAM model. Hsieh et~al.
\cite{HsiehHoHsuKo} presented an $O(n+m)$-time sequential
algorithm for the Hamiltonian problem on a distance-hereditary
graph and also proposed a parallel implementation of their
algorithm which solves the problem in $O(\log n)$ time using
$O((n+m)/\log n)$ processors on a PRAM model. A unified approach
to solving the Hamiltonian problems on distance-hereditary graphs
was presented in \cite{HungChangDH}, while Hsieh \cite{Hsieh}
presented an efficient parallel strategy for the 2HP problem on
the same class of graphs. Algorithms for the path cover problem on
other classes of graphs were proposed in \cite{HungChangCA, Nik,
SSSR93}.

\yy\noindent {\bf Road Map.} The paper is organized as follows. In
Section~2 we establish the notation and related terminology, and
we present background results. In Section~3 we describe our
algorithm for the 1PC problem, while in Section~4 we prove its
correctness and compute its time and space complexity. Section~5
presents some related results and in Section~6 we conclude the
paper and discuss possible future extensions.

\section{Theoretical Framework}

We consider finite undirected graphs with no loops or multiple
edges. For a graph~$G$, we denote its vertex and edge set by
$V(G)$ and $E(G)$, respectively. Let $S$ be a subset of the vertex
set of a graph~$G$. Then, the subgraph of $G$ induced by $S$ is
denoted by $G[S]$.

\vskip 0.2in 
\subsection{Structural Properties of Interval Graphs}

A graph $G$ is an {\it interval graph} if its vertices can be put
in a one-to-one correspondence with a family $F$ of intervals on
the real line such that two vertices are adjacent in $G$ if and
only if their corresponding intervals intersect. $F$ is called an
{\it intersection model} for $G$ \cite{AR90}. Interval graphs find
applications in genetics, molecular biology, archaeology, and
storage information retrieval \cite{Gol}. Interval graphs form an
important class of perfect graphs \cite{Gol} and many problems
that are NP-complete on arbitrary graphs are shown to admit
polynomial time algorithms on this class \cite{AR90, Gol, Keil}.
The class of interval graphs is {\it hereditary}, that is, every
induced subgraph of an interval graph $G$ is also an interval
graph. We state the following numbering for the vertices of an
interval graph proposed in \cite{RamRan}.

\medskip
\par\noindent
{\bf Lemma~2.1.} (Ramalingam and Rangan \cite{RamRan}): {\it The
vertices of any interval graph $G$ can be numbered with integers
$1, \ldots, |V(G)|$ such that if $i<j<k$ and $ik \in E(G)$ then
$jk \in E(G)$.}

\yyy\noindent As shown in \cite{RamRan}, the numbering of
Lemma~2.1, which results from numbering the intervals after
sorting them on their right ends \cite{AR90}, can be obtained in
linear time, that is, $O(m+n)$ time. An ordering of the vertices
according to this numbering is found to be quite useful in solving
many problems on interval graphs \cite{AR90, RamRan}. Throughout
the paper, the vertex numbered with $i$ will be denoted by $v_i$,
$1 \leq i \leq n$, and such an ordering will be denoted by $\pi$.
We say that $v_i<v_j$ if $i<j$, $1\leq i,j \leq n$.


\vskip 0.2in 
\subsection{Interval Graphs and the 1PC Problem}

Let $G$ be an interval graph with vertex set $V(G)$ and edge set
$E(G)$, $\mathcal{T}$ be a set containing a single vertex of
$V(G)$, and let $\mathcal{P_{\mathcal{T}}}(G)$ be a minimum
$1$-fixed-endpoint path cover of $G$ with respect to $\mathcal{T}$
of size $\lambda_\mathcal{T}(G)$ (or $\lambda_\mathcal{T}$ for
short); recall that the size of $\mathcal{P_{\mathcal{T}}}(G)$ is
the number of paths it contains. The vertex belonging to the set
$\mathcal{T}$ is called {\it terminal} vertex, and the set
$\mathcal{T}$ is called the {\it terminal set} of $G$, while those
of $V(G) - \mathcal{T}$ are called {\it non-terminal or free}
vertices. Thus, the set $\mathcal{P_{\mathcal{T}}}(G)$ contains
two types of paths, which we call {\it terminal} and {\it
non-terminal or free} paths: a {\it terminal path} $P_t$ is a path
having the terminal vertex as an endpoint and a {\it non-terminal
or free path} $P_f$ is a path having both its endpoints in $V(G) -
\mathcal{T}$. The set of the non-terminal paths in a minimum 1PC
of the graph $G$ is denoted by $N$, while $T$ denotes the set
containing the terminal path. Clearly, $|T|=1$ and
$\lambda_\mathcal{T}= |N| + 1$.

\begin{figure}[t]
\yy \hrule \yyy
  \centering
   \begin{minipage}[c]{2in}
    \centering
    \includegraphics[scale=1]{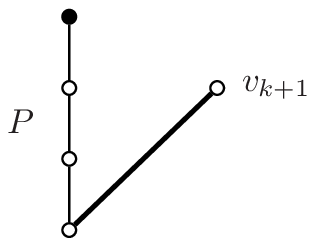}
    \center{\small{(a)}}
  \end{minipage}
  \begin{minipage}[c]{2in}
    \centering
    \includegraphics[scale=1]{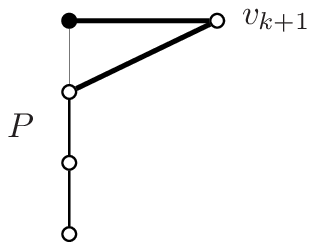}
   \center{\small{(b)}}
  \end{minipage}
  \begin{minipage}[c]{2in}
    \centering
    \includegraphics[scale=1]{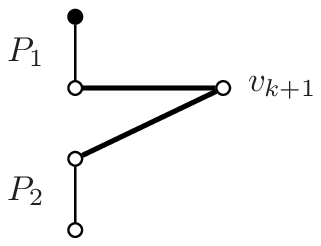}
   \center{\small{(c)}}
  \end{minipage}
  \caption{\small{Illustrating (a) connect, (b) insert, and (c) bridge operations; $P,P_1,P_2 \in
\mathcal{P_{\mathcal{T}}}(G[S])$.}} \yy \hrule \y\y
\label{operations}
\end{figure}

\y Our algorithm for computing a 1PC of an interval graph is based
on a greedy principle, visiting the vertices according to the
ordering $\pi=(v_1, v_2, \ldots, v_k, \ldots, v_n)$, and uses
three operations on the paths of a 1PC of $G[S]$, where $S=\{v_1,
v_2, \ldots, v_k\}$, $1 \leq k < n$. These three operations,
namely {\tt connect}, {\tt insert} and {\tt bridge} operations,
are described below and are illustrated in Fig.~\ref{operations}.

\begin{itemize}

\item [$\circ$]
{\tt Connect} operation: Let $v_i$ be a free endpoint of a path
$P$ of $\mathcal{P_{\mathcal{T}}}(G[S])$ and let $v_{k+1}$ be a
free or a terminal vertex such that $v_{k+1}$ sees $v_i$. We say
that we {\it connect} vertex $v_{k+1}$ to the path $P$, or,
equivalently, to the vertex $v_i$, if we extend the path $P$ by
adding an edge which joins vertex $v_{k+1}$ with vertex $v_i$.

%

\item [$\circ$] {\tt Insert} operation: Let $P=(\ldots, v_i, v_j, \ldots)$, $i \neq j$, $i,j \in [1,k]$, be a
path of $\mathcal{P_{\mathcal{T}}}(G[S])$ and let $v_{k+1}$ be a
free vertex such that $v_{k+1}$ sees $v_i$ and $v_j$. We say that
we {\it insert} vertex $v_{k+1}$ into $P$, if we replace the path
$P$ with the path $P'=(\ldots, v_i, v_{k+1}, v_j, \ldots)$.

\item [$\circ$] {\tt Bridge} operation: Let $P_1$ and $P_2$ be two
paths of $\mathcal{P_{\mathcal{T}}}(G[S])$ and let $v_{k+1}$ be a
free vertex. We say that we {\it bridge} the two paths $P_1$ and
$P_2$ using vertex $v_{k+1}$ if we connect $v_{k+1}$ with a free
endpoint of $P_1$ and a free endpoint of $P_2$.

\end{itemize}

Let $P$ be a path of $\mathcal{P_{\mathcal{T}}}(G)$ and let $v_i$
and $v_j$ be its endpoints. We say that $v_i$ is the left (resp.
right) endpoint of the path and $v_j$ is the right (resp. left)
endpoint of the path if $v_i < v_j$ (resp. $v_j < v_i$).
Throughout the paper, a trivial path (i.e. a path consisting of
one vertex) is considered to have two endpoints, while a trivial
path consisting of the terminal vertex $u \in \mathcal{T}$ is
considered to have one terminal endpoint and one free endpoint.

\y Let $G$ be an interval graph on $n$ vertices and let
$\mathcal{P_{\mathcal{T}}}(G)$ be a minimum 1PC of size
$\lambda_\mathcal{T}$. Since a trivial path is considered to have
two endpoints, the number of endpoints in
$\mathcal{P_{\mathcal{T}}}(G)$ is $2\lambda_\mathcal{T}$. For each
vertex $v_i$ we denote by $d(v_i)$ the number of neighbors of
$v_i$ in $\mathcal{P_{\mathcal{T}}}(G)$; that is, $d(v_i) \in
\{0,1,2\}$. We call {\it d-connectivity} of
$\mathcal{P_{\mathcal{T}}}(G)$ the sum of $d(v_1), d(v_2), \ldots,
d(v_n)$. It is easy to see that $\sum^n_{i=1} d(v_i)=
2(n-\lambda_\mathcal{T})$. Clearly, any minimum 1PC
$\mathcal{P_{\mathcal{T}}}(G)$ has d-connectivity equal to
$2(n-\lambda_\mathcal{T})$.

\vskip 0.3in 
\section{The Algorithm}

We next present an algorithm for the 1PC problem on interval
graphs. Our algorithm takes as input an interval graph $G$ on $n$
vertices and $m$ edges and a set $\mathcal{T}=\{u\}$ containing
the terminal vertex $u \in V(G)$, and computes a minimum 1PC of
$G$ in $O(n^2)$ time; it is based on a greedy principle to extend
a path of a minimum 1PC using operations on the left and right
endpoints of its paths and properties of the graph $G[\{v_1, v_2,
\ldots, v_i\}-\{u\}]$, $1 \leq i \leq n$. We point out that, if a
vertex sees the two endpoints of only one non-terminal path $P$,
it is connected to the left endpoint of the path $P$. Furthermore,
for each vertex $v_i$, $1 \leq i < j$, we denote by
$\varepsilon^{(j)}_i$ the number of endpoints $v_\kappa$ belonging
to different paths with index $\kappa \in (i,j]$. We also define
$\varepsilon^{(i)}_i=0$ and
$\varepsilon^{(i)}_0=\lambda_\mathcal{T}(G[v_1, \ldots, v_i])$, $1
\leq i \leq n$. The algorithm works as follows:

\bigskip \noindent{\it Algorithm Minimum\_1PC} \y \noindent{\it
Input:} an interval graph $G$ on $n$ vertices and $m$ edges and a
vertex $u \in V(G)$; \y \noindent{\it Output:} a minimum 1PC
$\mathcal{P_{\mathcal{T}}}(G)$ of the interval graph $G$; \y

\begin{enumerate}
  \item Construct the ordering $\pi$ of the vertices of $G$;
  \item Execute the subroutine $process(\pi)$; the minimum
  1PC $\mathcal{P_{\mathcal{T}}}(G)$ is the set
  of paths returned by the subroutine;
\end{enumerate}

\y \noindent where the description of the subroutine
$process(\pi)$ is presented below.

\yyy It is easy to see that, if $\lambda_\mathcal{T}(G)$ is the
size of a minimum 1PC of $G$ with respect to $\mathcal{T}=\{v_t\}$
then the size of a minimum 1PC of $G-\{v_t\}$ is either
$\lambda_\mathcal{T}(G)$ or $\lambda_\mathcal{T}(G)-1$. Indeed,
suppose that the size of a minimum 1PC of $G-\{v_t\}$ is
$\lambda_\mathcal{T}(G)+1$. Since a terminal vertex cannot
decrease the size of a minimum 1PC, we have
$\lambda_\mathcal{T}(G) \geq \lambda_\mathcal{T}(G-\{v_t\})$.
Thus, $\lambda_\mathcal{T}(G) \geq \lambda_\mathcal{T}(G)+1$, a
contradiction. Suppose now that the size of a minimum 1PC
$\mathcal{P_{\mathcal{T}}}(G-\{v_t\})$ of $G-\{v_t\}$ is
$\lambda_\mathcal{T}(G)-2$. Then, adding a trivial path containing
vertex $v_t$ to $\mathcal{P_{\mathcal{T}}}(G-\{v_t\})$ results to
a 1PC of $G$ of size $\lambda_\mathcal{T}(G)-1$, a contradiction.

\y We next describe the operation {\tt bridge} in detail. Note
that in most cases we bridge two paths through their leftmost free
endpoints. Suppose that when vertex $v_i$ is processed it sees at
least one free endpoint of a non-terminal path $P_1$, say, $v_j$,
and at least the free endpoint of the terminal path $P_2$, say,
$v_\ell$, and both endpoints of a non-terminal path $P_3$, say,
$v_r$ and $v_s$. Let $v_z$, $v_t$ and $v_r$ be the left endpoints
and $v_j$, $v_\ell$ and $v_s$ be the right endpoints of the paths
$P_1$, $P_2$ and $P_3$, respectively. There exist three cases
where we do not bridge two paths through the leftmost free
endpoints (see Fig.~\ref{bridgeOP}(a)-(c)). In these three cases
the bridge operation works as follows:

\yy \noindent (a) $v_z<v_j<v_t<v_\ell<v_r<v_s$: we bridge $P_1$
and $P_3$ through $v_z$ (or, $v_j$ if $v_z \notin N(v_i)$) and
$v_r$.

\bigskip
\small{ \yy \hrule \y\y\y \noindent{\it process }($\pi$) \y
\noindent{\it Input:} the ordering $\pi$ of the vertices of $G$
and the index $t$ of the terminal vertex $u$;\y \noindent{\it
Output:} a minimum 1PC $\mathcal{P_{\mathcal{T}}}$ of $G$;
\y\noindent
  $\lambda_\mathcal{T}=1$;
  $P_{\lambda_\mathcal{T}}=(v_1)$; $\varepsilon^{(1)}_0=1$;\\
  $p^\ell_1=v_1$; $p^r_1=v_1$; \phantom{\tt else}{\it \{the left and right endpoints of the path $P_{\lambda_\mathcal{T}}$\}}\\
{\tt for} $i=2$ {\tt to} $n$ {\tt do}\\
$\rhd$ {\tt if} $i \neq t$ {\tt then}\\
\phantom{\tt el}$\circ$ {\tt if} $N(v_i) \neq \emptyset$ {\tt and} $\varepsilon^{(i-1)}_{j-1} \geq 2$ {\tt then}\phantom{\tt else}{\it \{$v_j$ is the leftmost neighbor of $v_i$ in $G[\{v_1, \ldots, v_{i-1}\}]$\}}\\
\phantom{\tt el}\phantom{\tt el}\phantom{\tt el}{\tt if} at
least two endpoints are free vertices {\tt then} {\tt bridge}; $\lambda_\mathcal{T}=\lambda_\mathcal{T}-1$;\\
 \phantom{\tt
el}\phantom{\tt el}\phantom{\tt el}{\tt else}\phantom{\tt else} {\it \{$\varepsilon^{(i-1)}_{j-1}=2$ and one endpoint is the terminal vertex $v_t$, and the other, say, $v_f$, is a free vertex.\}}\\
\phantom{\tt el}\phantom{\tt el}\phantom{\tt el}\phantom{\tt
el}{\tt if} process($\{v_1, \ldots, v_{i-1}\}-\{v_t\}$) returns a
1PC $\mathcal{P_{\mathcal{T}}}(G[\{v_1, \ldots,
v_{i-1}\}-\{v_t\}])$ of $\lambda_\mathcal{T}-1$
paths {\tt then}\\
\phantom{\tt el}\phantom{\tt el}\phantom{\tt el}\phantom{\tt
el}\phantom{\tt el} {\tt connect} $v_i$ to the leftmost endpoint
of $\mathcal{P_{\mathcal{T}}}(G[\{v_1, \ldots,
v_{i-1}\}-\{v_t\}])$; {\tt connect} $v_t$ to $v_i$;\\
\phantom{\tt el}\phantom{\tt el}\phantom{\tt el}\phantom{\tt
el}\phantom{\tt el} $\lambda_\mathcal{T}=\lambda_\mathcal{T}-1$;\\
\phantom{\tt el}\phantom{\tt el}\phantom{\tt el}\phantom{\tt
el}{\tt else-if} process($\{v_1, \ldots, v_{i-1}\}-\{v_t\}$)
returns a 1PC $\mathcal{P_{\mathcal{T}}}(G[\{v_1, \ldots,
v_{i-1}\}-\{v_t\}])$ of $\lambda_\mathcal{T}$
paths {\tt then}\\
\phantom{\tt el}\phantom{\tt el}\phantom{\tt el}\phantom{\tt
el}\phantom{\tt el}{\tt if} $v_t$ can be connected to the leftmost left endpoint and then $v_i$ can bridge paths {\tt then}\\
\phantom{\tt el}\phantom{\tt el}\phantom{\tt el}\phantom{\tt
el}\phantom{\tt el}\phantom{\tt el} {\tt connect} $v_t$ to the
leftmost left endpoint of $\mathcal{P_{\mathcal{T}}}(G[\{v_1,
\ldots,
v_{i-1}\}-\{v_t\}])$; {\tt bridge}; \\
\phantom{\tt el}\phantom{\tt el}\phantom{\tt el}\phantom{\tt
el}\phantom{\tt el}\phantom{\tt el} $\lambda_\mathcal{T}=\lambda_\mathcal{T}-1$;\\
\phantom{\tt el}\phantom{\tt el}\phantom{\tt el}\phantom{\tt
el}{\tt else} {\tt connect} $v_i$ to the leftmost endpoint;\\
\phantom{\tt el}$\circ$ {\tt if} $N(v_i) \neq \emptyset$ {\tt and}
$\varepsilon^{(i-1)}_{j-1} =1$ {\tt and}
the endpoint $v_f, j \leq f \leq i-1$, is a free vertex {\tt then}\\
\phantom{\tt el}\phantom{\tt el}\phantom{\tt el}{\tt if} $v_i$ sees an internal vertex $v_s$ {\tt then} {\tt connect\_break};\\
\phantom{\tt el}\phantom{\tt el}\phantom{\tt el}{\tt else connect} $v_i$ to the leftmost free endpoint;\\
\phantom{\tt el}$\circ$ {\tt if} $N(v_i)=\emptyset$ {\tt or}
$\varepsilon^{(i-1)}_{j-1} =0$ {\tt or}
($\varepsilon^{(i-1)}_{j-1} =1$ {\tt and}
the endpoint $v_t, j \leq t \leq i-1$, is the terminal vertex) {\tt then}\\
\phantom{\tt el}\phantom{\tt el}\phantom{\tt el}{\tt if} $v_i$ has
two consecutive neighbors into a path {\tt then} {\tt insert} $v_i$ into the path;\\
\phantom{\tt el}\phantom{\tt el}\phantom{\tt el}{\tt else-if}
$v_i$ sees an internal vertex $v_s$ {\tt then}\\
\phantom{\tt el}\phantom{\tt el}\phantom{\tt el}\phantom{\tt
el}{\tt if} $v_sv_a$ is an edge of a path $P_k$ and $v_a$ sees an endpoint $v_b$ of a path $P_{k'}, k \neq k'$ {\tt then} \\
\phantom{\tt el}\phantom{\tt el}\phantom{\tt el}\phantom{\tt
el}\phantom{\tt
el} remove the edge $v_sv_a$ of $P$; {\tt connect} $v_a$ to $v_b$; {\tt connect} $v_i$ to $v_j$;\\
\phantom{\tt el}\phantom{\tt el}\phantom{\tt el}\phantom{\tt
el}{\tt else} {\tt new\_path}; $\lambda_\mathcal{T}=\lambda_\mathcal{T}+1$;\\
\phantom{\tt el}\phantom{\tt el}\phantom{\tt el}{\tt else}
$\lambda_\mathcal{T}=\lambda_\mathcal{T}+1$;
$P_{\lambda_\mathcal{T}}=(v_i)$;\\
\phantom{\tt el}{\tt endif};\\
$\rhd$ {\tt if} $i=t$ {\tt then}\\
\phantom{\tt el}\phantom{\tt el}{\tt if}
$\varepsilon^{(i-1)}_{j-1} \geq 1$ {\tt then} {\tt connect} $v_i$
to the leftmost endpoint of $\mathcal{P_{\mathcal{T}}}(G[\{v_1,
\ldots, v_{i-1}\}])$;\\
\phantom{\tt el}\phantom{\tt el}{\tt else}
$\lambda_\mathcal{T}=\lambda_\mathcal{T}+1$;
$P_{\lambda_\mathcal{T}}=(v_i)$;\\
\phantom{\tt el}{\tt endif};\\
\phantom{\tt el}{\tt for} $k=0$ {\tt to} $i$ {\tt do} {\tt update} $\varepsilon^{(i)}_k$;   {\tt update\_endpoints};\\
{\tt endfor};\\
  $\mathcal{P_{\mathcal{T}}}(G)=\{P_1, \ldots,
  P_{\lambda_\mathcal{T}}\}$.
 \yyy \hrule \y\y}
 \normalsize{
\bigskip

\begin{figure}[t]
\hrule \bigskip \hspace*{0.3in}
   \begin{minipage}[l]{2in}
    \includegraphics[scale=0.8]{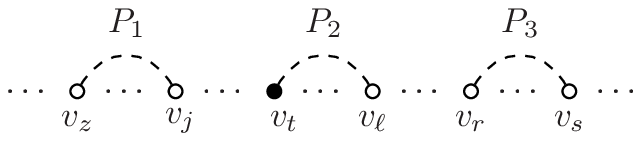}
    \hspace*{1in}\small{(a)}
  \end{minipage}
  \begin{minipage}[r]{2in}
   \hspace*{1.1in}
    \includegraphics[scale=0.8]{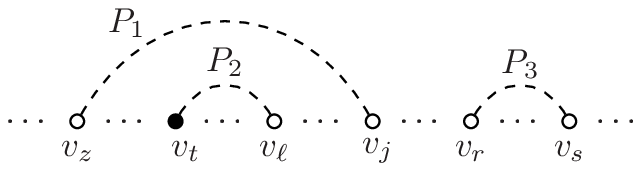}
   \hspace*{2.2in}\small{(b)}
  \end{minipage}
\vspace{0.2in}

\hspace*{0.25in}
  \begin{minipage}[c]{2in}
    \hspace*{0.05in}
    \includegraphics[scale=0.8]{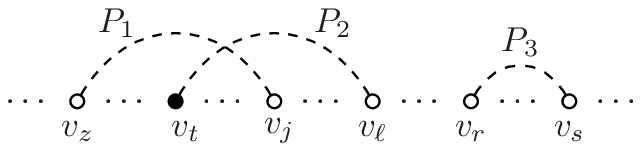}
   \hspace*{1.05in}\small{(c)}
  \end{minipage}
 \begin{minipage}[c]{2in}
    \hspace*{0.9in}
    \includegraphics[scale=0.8]{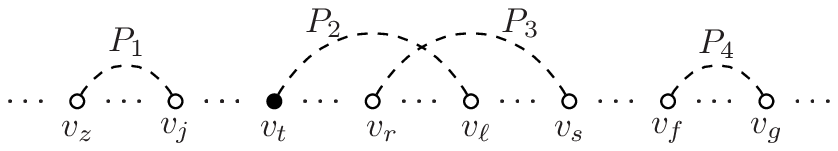}
   \hspace*{2.25in}\small{(d)}
  \end{minipage}
\vspace{0.2in}
 \vspace*{-0.2in}

\hspace*{0.25in}
 \begin{minipage}[c]{2in}
    \hspace*{1.5in}
    \includegraphics[scale=0.8]{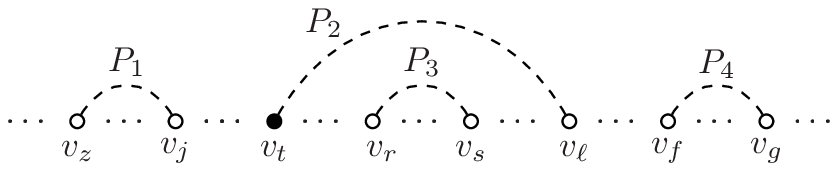}
   \hspace*{2.9in}\small{(e)}
  \end{minipage}

    \caption{\small{Illustrating some cases of the bridge operation.}}
\bigskip \hrule
 \label{bridgeOP}
\end{figure}

\y\noindent (b) $v_z<v_t<v_\ell<v_j<v_r<v_s$: if $v_z \notin
N(v_i)$ we bridge $P_2$ and $P_3$ through $v_\ell$ and $v_r$;
otherwise, we \phantom{(b)} bridge $P_1$ and $P_2$ through $v_z$
and $v_\ell$.

\y \noindent (c) $v_z<v_t<v_j<v_\ell<v_r<v_s$: we bridge $P_1$ and
$P_3$ through $v_z$ (or, $v_j$ if $v_z \notin N(v_i)$) and $v_r$.

\yy Suppose now that $P_1$ is a non-terminal path having $v_z$ and
$v_j$ as its left and right endpoints, respectively, $P_2$ is the
terminal path with left endpoint $v_t$ and right endpoint
$v_\ell$. Also, let $P_3$ be a non-terminal path with left and
right endpoints $v_r$ and $v_s$, respectively, and $P_4$ a
non-terminal path with left and right endpoints $v_f$ and $v_g$,
respectively (see Fig.~\ref{bridgeOP}(d)-(e)). We distinguish the
following two cases:

\y \noindent (d) $v_z<v_j<v_t<v_r<v_\ell<v_s<v_f<v_g$: if $v_z \in
N(v_i)$ or $v_j \in N(v_i)$ we bridge $P_1$ and $P_3$ through
\phantom{(d)} $v_z$ ($v_j$ if $v_z \notin N(v_i)$) and $v_r$. If
$v_\ell \in N(v_i)$ and $v_r \notin N(v_i)$ we bridge $P_2$ and
$P_4$ through $v_\ell$ and $v_f$.

\y \noindent (e) $v_z<v_j<v_t<v_r<v_s<v_\ell<v_f<v_g$: if $v_z \in
N(v_i)$ or $v_j \in N(v_i)$ we bridge $P_1$ and $P_3$ through
\phantom{(e)} $v_i$ ($v_j$ if $v_z \notin N(v_i)$) and $v_r$;
otherwise, we bridge $P_3$ and $P_4$ through $v_r$ ($v_s$ if $v_r
\notin N(v_i)$) and $v_f$.


\yy \noindent Figure~\ref{bridgeOP} presents cases (a)-(e).
Suppose that we have the two paths $P_2$ and $P_3$ of case~(e) and
vertex $v_i$ sees both $v_r$ and $v_s$, that is, $P_2=(v_t,
\ldots, v_a, v_b, v_c, \ldots, v_\ell)$ and $P_3=(v_r, \ldots,
v_s)$, where $v_a<v_s<v_b$ and $v_s<v_c$. Then, the bridge
operation constructs the path $P=(v_t, \ldots, v_a, v_b, v_s,
\ldots, v_r, v_i, v_c, \ldots, v_\ell)$. Suppose now that we have
the two paths $P_1$ and $P_2$ of case~(c) and vertex $v_i$ sees
all vertices with index greater or equal to $z$, that is,
$P_1=(v_z, \ldots, v_j)$ and $P_2=(v_t, \ldots, v_a, v_b, v_c,
\ldots, v_\ell)$, where $v_a<v_j<v_b$ and $v_j<v_c$. Then, the
bridge operation constructs the path $P=(v_t, \ldots, v_a, v_b,
v_j, \ldots, v_z, v_i, v_c, \ldots, v_\ell)$. Suppose that there
exist two paths $P_2$ and $P_3$ as in case~(d) and vertex $v_i$
sees all vertices with index $k$, $d \leq k$, where $r<d \leq
\ell$, that is, $P_2=(v_t, \ldots, v_a, v_b, v_c, \ldots, v_\ell)$
and $P_3=(v_r, \ldots, v_s)$, where $v_a<v_r<v_b$ and $v_r<v_c$.
If $d<c$ then the bridge operation constructs the path $P=(v_t,
\ldots, v_a, v_b, v_r, \ldots, v_s, v_i, v_c, \ldots, v_\ell)$;
otherwise, it constructs the path $P=(v_t, \ldots, v_a, v_b, v_r,
\ldots, v_s, v_i,$ $v_\ell, \ldots, v_c)$. If there exist two
paths $P_1$ and $P_2$ as in case~(b) and vertex $v_i$ sees all
vertices with index greater or equal to $z$, that is, $P_1=(v_z,
\ldots, v_a, v_b, v_c, \ldots, v_j)$ and $P_2=(v_t, \ldots,
v_\ell)$, where $v_a<v_\ell<v_b$ and $v_\ell<v_c$, then the bridge
operation constructs the path $P=(v_t, \ldots, v_\ell, v_b, v_a,
\ldots, v_z, v_i, v_c, \ldots, v_j)$, if $c<j$, or the path
$P=(v_t, \ldots, v_\ell, v_b, v_a, \ldots, v_z, v_i, v_j, \ldots,
v_c)$, if $j<c$.

\y We next describe the operation {\tt new\_path} which creates a
new path when the vertex $v_i$ is processed. There exist three
cases where operation new\_path creates a new non-trivial path
while in all other cases it creates a new trivial path. Suppose
that $v_i$ sees an internal vertex $v_j$ belonging to a path
$P=(v_s, \ldots, v_r, v_j, v_\ell, \ldots, v_t)$ such that
$v_r<v_s<v_t<v_\ell<v_j$. We remove the edge $v_jv_\ell$ from $P$
and we obtain $P_1=(v_s, \ldots, v_r, v_j)$ and $P_2=(v_t, \ldots,
v_\ell)$. Then, we connect $v_i$ to $v_j$. The case where $v_jv_s
\in E(G)$ and $v_jv_r \notin E(G)$ is similar. If $v_i$ sees an
internal vertex $v_j$ belonging to a path $P=(v_r, \ldots, v_s,
v_j, v_\ell, \ldots, v_t)$ such that $v_r<v_t<v_\ell<v_s<v_j$, we
remove the edge $v_sv_j$ from $P$ and we obtain $P_1=(v_r, \ldots,
v_s)$ and $P_2=(v_t, \ldots, v_\ell, v_j)$. Then, we connect $v_i$
to $v_j$. Suppose now that $v_i$ sees an internal vertex $v_j$
belonging to a path $P=(v_\ell, \ldots, v_s, v_j, v_r, \ldots,
v_t)$ such that $v_\ell<v_t<v_s<v_r<v_j$. We remove the edge
$v_jv_r$ from $P$ and we obtain $P_1=(v_\ell, \ldots, v_s, v_j)$
and $P_2=(v_t, \ldots, v_r)$. Then, we connect $v_i$ to $v_j$. The
above cases, where the operation new\_path creates a new
non-trivial path, are described below:

\y \noindent (a) $v_r<v_s<v_t<v_\ell<v_j$: We create paths
$P_1=(v_s, \ldots, v_r, v_j, v_i)$ and $P_2=(v_t, \ldots,
v_\ell)$. The case \phantom{(a) }where $v_jv_s \in E(G)$ and
$v_jv_r \notin E(G)$ is similar.

\y \noindent (b) $v_r<v_t<v_\ell<v_s<v_j$: We create paths
$P_1=(v_r, \ldots, v_s)$ and $P_2=(v_t, \ldots, v_\ell, v_j,
v_i)$.

\y \noindent (c)\phantom{c}$v_\ell<v_t<v_s<v_r<v_j$: We create
paths $P_1=(v_\ell, \ldots, v_s, v_j, v_i)$ and $P_2=(v_t, \ldots,
v_r)$.


\yy Note that, the rightmost endpoint of a path in
$\mathcal{P_{\mathcal{T}}}(G[\{v_1, \ldots, v_{i-1}\}])$ is $v_t$
and, thus, $\varepsilon^{(i-1)}_t=0$. The 1PC
$\mathcal{P_{\mathcal{T}}}(G)$ of the graph $G$ in each of the
above cases contains two new endpoints, vertices $v_i$ and
$v_{k'}$ such that $t<k'$; thus, $\varepsilon^{(i)}_t=2$.
Figure~\ref{newOP} presents the above cases.

\y The operation {\tt connect\_break} is similar to the operation
new\_path. Specifically, suppose that in the above cases (a)-(c)
there exists a path $P=(v_a, \ldots, v_b)$ such that
$v_j<v_a<v_b<v_i$. Then, the operation connect\_break works
similarly to the operation new\_path; the only difference is that
$v_i$ is also connected to $v_a$. Note that, in all other cases,
operation connect\_break coincides with the operation connect.

\begin{figure}[t]
\hrule \bigskip \hspace*{0.6in}
   \begin{minipage}[l]{2in}
\includegraphics[scale=0.8]{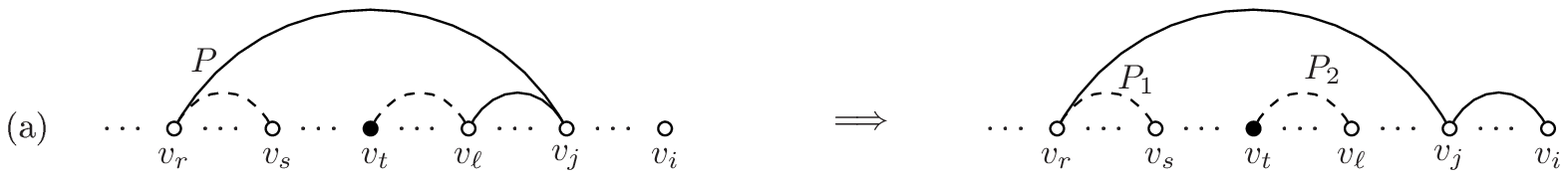}
  \end{minipage}
\vspace{0.2in}

\hspace*{0.6in}
  \begin{minipage}[r]{2in}
    \includegraphics[scale=0.8]{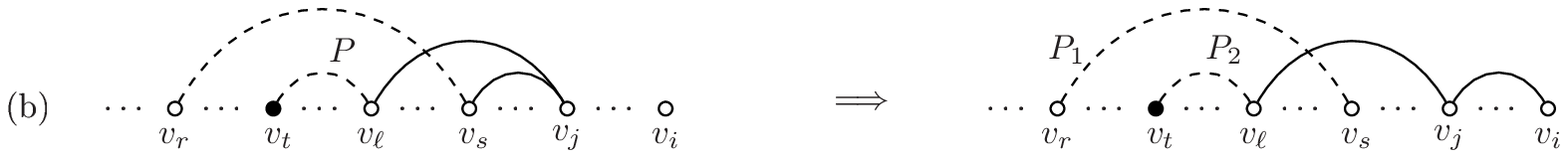}
  \end{minipage}
\vspace{0.2in}

\hspace*{0.6in}
  \begin{minipage}[c]{2in}
   \includegraphics[scale=0.8]{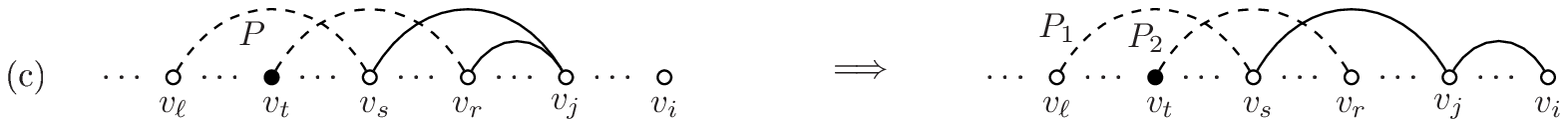}
  \end{minipage}

    \y\caption{\small{Illustrating some cases of the new\_path operation.}}
\bigskip \hrule
 \label{newOP}
\end{figure}

\y Concerning the ordering of the endpoints of the paths of the
1PC constructed by Algorithm Minimum\_1PC, we prove the following
lemmata.

\medskip
\par\noindent
{\bf Lemma~3.1.} {\it Let $G$ be an interval graph with no
terminal vertex. Let $P_s$, $1 \leq s \leq \lambda_\mathcal{T}$,
be a path in the 1PC $\mathcal{P_{\mathcal{T}}}(G)$ of the graph
$G$ constructed by Algorithm Minimum\_1PC and let $v_i$ and $v_j$
be the left and right endpoints of $P_s$, respectively. Then,
there is no path $P_t \in \mathcal{P_{\mathcal{T}}}(G)$, $1 \leq t
\leq \lambda_\mathcal{T}$, $t\neq s$, such that $v_i<v_t<v_j$ or
$v_i<v_\ell<v_j$, where $v_k$ and $v_\ell$ are the left and right
endpoints of $P_t$, respectively.}

\yy \noindent {\sl Proof.} \s Let $P_s$, $1 \leq s \leq
\lambda_\mathcal{T}$ be a path in the 1PC
$\mathcal{P_{\mathcal{T}}}(G)$ constructed by Algorithm
Minimum\_1PC and let $v_i$ and $v_j$ be its left and right
endpoints, respectively. Let $P_t \in
\mathcal{P_{\mathcal{T}}}(G)$, $1 \leq t \leq
\lambda_\mathcal{T}$, $t\neq s$, and let $v_k$ and $v_\ell$ be its
left and right endpoints, respectively. Suppose that
$v_i<v_k<v_j$. Since $v_i$ and $v_j$ are the endpoints of $P_s$
and $v_i<v_k<v_j$, the path $P_s$ contains at least one edge, say,
$v_av_b$, such that $v_a<v_k<v_b$. Clearly, vertices $v_a$ and
$v_b$ are free vertices. Since $v_av_b \in E(G)$, we also have
$v_kv_b \in E(G)$. Then, according to Algorithm Minimum\_1PC, when
vertex $v_b$ is processed, vertex $v_k$ is an endpoint of the path
$P_t$, and, thus, $v_b$ bridges the paths $P_s$ and $P_t$ through
vertices $v_a$ and $v_k$, a contradiction. Similarly, we can prove
that $v_\ell<v_i$ or $v_\ell>v_j$. \ \qed

\y Using similar arguments we can prove the following lemma:

\medskip
\par\noindent
{\bf Lemma~3.2.} {\it Let $G$ be an interval graph containing a
terminal vertex. Let $P_s$, $1 \leq s \leq \lambda_\mathcal{T}$,
be a non-terminal path in the 1PC $\mathcal{P_{\mathcal{T}}}(G)$
of the graph $G$ constructed by Algorithm Minimum\_1PC and let
$v_i$ and $v_j$ be the left and right endpoints of $P_s$,
respectively. Then, there is no non-terminal path $P_t \in
\mathcal{P_{\mathcal{T}}}(G)$, $1 \leq t \leq
\lambda_\mathcal{T}$, $t\neq s$, such that $v_i<v_k<v_j$ or
$v_i<v_\ell<v_j$, where $v_k$ and $v_\ell$ are the left and right
endpoints of $P_t$, respectively.}

\section{Correctness and Time Complexity}

Let $G$ be an interval graph on $n$~vertices and~$m$ edges and let
$\mathcal{T}$ be a subset of $V(G)$ containing a single vertex. In
order to prove the correctness of Algorithm Minimum\_1PC, we use
induction on $n$. We also prove a property of the minimum 1PC
$\mathcal{P}_\mathcal{T}(G)$ of $G$ constructed by our algorithm:
Algorithm Minimum\_1PC computes a minimum 1PC
$\mathcal{P_{\mathcal{T}}}(G)$ of the graph $G$ having
$\varepsilon^{(n)}_i$ endpoints $v_\kappa$ belonging to different
paths with index $\kappa \in (i,n]$, $1 \leq i \leq n$, such that
there is no other minimum 1PC $\mathcal{P'_{\mathcal{T}}}(G)$
having $\varepsilon'^{(n)}_i$ endpoints $v_{\kappa'}$ belonging to
different paths with index $\kappa' \in (i,n]$ such that
$\varepsilon'^{(n)}_i>\varepsilon^{(n)}_i$, $1 \leq i < \varrho$,
where $\varrho$ is the index of the rightmost endpoint of a path
in $\mathcal{P_{\mathcal{T}}}(G)$. Furthermore, one of the
following holds:

\noindent (i) $\varepsilon'^{(n)}_i \leq \varepsilon^{(n)}_i$,
$\varrho \leq i \leq n$, or

\noindent (ii) if $\varepsilon'^{(n)}_i = \varepsilon^{(n)}_i +1$,
$\varrho \leq i < \varrho'$ and $\varepsilon'^{(n)}_i =
\varepsilon^{(n)}_i$, $\varrho' \leq i \leq n$, where $\varrho'$
is the index of the rightmost endpoint of a path in
$\mathcal{P'_{\mathcal{T}}}(G)$, then there exists a vertex $v_z$
such that $\varepsilon^{(n)}_z > \varepsilon'^{(n)}_z$, $1 \leq z
< \varrho$ and there exists no vertex $v_{z'}$, $1 \leq z'
<\varrho$, such that
$\varepsilon'^{(n)}_{z'}>\varepsilon^{(n)}_{z'}$.

\noindent Recall that a trivial path has two endpoints that
coincide. Hence, we prove the following theorem.

\bigskip
\par\noindent
{\bf Theorem~4.1.} {\it Let $G$ be an interval graph on
$n$~vertices and~$m$ edges and let $v \in V(G)$. Algorithm
Minimum\_1PC computes a minimum 1PC $\mathcal{P_{\mathcal{T}}}(G)$
of the graph $G$ having $\varepsilon^{(n)}_i$ endpoints $v_\kappa$
belonging to different paths with index $\kappa \in (i,n]$, $1
\leq i \leq n$, such that there is no other minimum 1PC
$\mathcal{P'_{\mathcal{T}}}(G)$ having $\varepsilon'^{(n)}_i$
endpoints $v_{\kappa'}$ belonging to different paths with index
$\kappa' \in (i,n]$ such that
$\varepsilon'^{(n)}_i>\varepsilon^{(n)}_i$, $1 \leq i < \varrho$,
where $\varrho$ is the index of the rightmost endpoint of a path
in $\mathcal{P_{\mathcal{T}}}(G)$. Furthermore, one of the
following holds:
\begin{itemize}
\item[(i)] $\varepsilon'^{(n)}_i \leq \varepsilon^{(n)}_i$,
$\varrho \leq i \leq n$, or \item[(ii)] if $\varepsilon'^{(n)}_i =
\varepsilon^{(n)}_i +1$, $\varrho \leq i < \varrho'$ and
$\varepsilon'^{(n)}_i = \varepsilon^{(n)}_i$, $\varrho' \leq i
\leq n$, where $\varrho'$ is the index of the rightmost endpoint
of a path in $\mathcal{P'_{\mathcal{T}}}(G)$, then there exists a
vertex $v_z$ such that $\varepsilon^{(n)}_z >
\varepsilon'^{(n)}_z$, $1 \leq z < \varrho$ and there exists no
vertex $v_{z'}$, $1 \leq z' <\varrho$, such that
$\varepsilon'^{(n)}_{z'}>\varepsilon^{(n)}_{z'}$.
\end{itemize}}

\yy \noindent {\sl Proof.} \s We use induction on $n$. The basis
$n=1$ is trivial. Assume that Algorithm Minimum\_1PC computes a
minimum 1PC $\mathcal{P_{\mathcal{T}}}(G[S])$ of every interval
graph $G[S]$, $S=\{v_1, v_2, \ldots, v_{n-1}\}$, with at most
$n-1$ vertices having $\varepsilon^{(n-1)}_i$ endpoints $v_\kappa$
belonging to different paths with index $\kappa \in (i,n-1]$, $1
\leq i \leq n-1$, such that there is no other minimum 1PC
$\mathcal{P'_{\mathcal{T}}}(G[S])$ having $\varepsilon'^{(n-1)}_i$
endpoints $v_{\kappa'}$ belonging to different paths with index
$\kappa' \in (i,n-1]$ such that
$\varepsilon'^{(n-1)}_i>\varepsilon^{(n-1)}_i$, $1 \leq i < d$,
where $d$ is the index of the rightmost endpoint of a path in
$\mathcal{P_{\mathcal{T}}}(G[S])$. Furthermore, one of the
following holds:
\begin{itemize}
\item[(i)] $\varepsilon'^{(n-1)}_i \leq \varepsilon^{(n-1)}_i$, $d
\leq i \leq n-1$, or

\item[(ii)] if $\varepsilon'^{(n-1)}_i = \varepsilon^{(n-1)}_i
+1$, $d \leq i < d'$ and $\varepsilon'^{(n-1)}_i =
\varepsilon^{(n-1)}_i$, $d' \leq i \leq n-1$, where $d'$ is the
index of the rightmost endpoint of a path in
$\mathcal{P'_{\mathcal{T}}}(G[S])$, then there exists a vertex
$v_q$ such that $\varepsilon^{(n-1)}_q > \varepsilon'^{(n-1)}_q$,
$1 \leq q < d$ and there exists no vertex $v_{q'}$, $1 \leq q'
<d$, such that
$\varepsilon'^{(n-1)}_{q'}>\varepsilon^{(n-1)}_{q'}$.

\end{itemize}

\noindent Let $\lambda_\mathcal{T}(G[S])$ be the size of
$\mathcal{P_{\mathcal{T}}}(G[S])$. We show that the algorithm
computes a minimum 1PC $\mathcal{P_{\mathcal{T}}}(G)$ of the graph
$G$ having $\varepsilon^{(n)}_i$ endpoints $v_\kappa$ belonging to
different paths with index $\kappa \in (i,n]$, $1 \leq i \leq n$,
such that there is no other minimum 1PC
$\mathcal{P'_{\mathcal{T}}}(G)$ having $\varepsilon'^{(n)}_i$
endpoints $v_{\kappa'}$ belonging to different paths with index
$\kappa' \in (i,n]$ such that
$\varepsilon'^{(n)}_i>\varepsilon^{(n)}_i$, $1 \leq i < \varrho$,
where $\varrho$ is the index of the rightmost endpoint of a path
in $\mathcal{P_{\mathcal{T}}}(G)$. Furthermore, one of the
following holds:
\begin{itemize}
\item[(i)] $\varepsilon'^{(n)}_i \leq \varepsilon^{(n)}_i$,
$\varrho \leq i \leq n$, or \item[(ii)] if $\varepsilon'^{(n)}_i =
\varepsilon^{(n)}_i +1$, $\varrho \leq i < \varrho'$ and
$\varepsilon'^{(n)}_i = \varepsilon^{(n)}_i$, $\varrho' \leq i
\leq n$, where $\varrho'$ is the index of the rightmost endpoint
of a path in $\mathcal{P'_{\mathcal{T}}}(G)$, then there exists a
vertex $v_z$ such that $\varepsilon^{(n)}_z >
\varepsilon'^{(n)}_z$, $1 \leq z < \varrho$ and there exists no
vertex $v_{z'}$, $1 \leq z' <\varrho$, such that
$\varepsilon'^{(n)}_{z'}>\varepsilon^{(n)}_{z'}$.

\end{itemize}

\yy {\bf Case~A:} Vertex $v_n$ is not the terminal vertex. Let
$\lambda_\mathcal{T}(G)$ be the size of
$\mathcal{P_{\mathcal{T}}}(G)$. Clearly, the size
$\lambda'_\mathcal{T}(G)$ of a minimum 1PC of $G$ is equal to
$\lambda_\mathcal{T}(G[S])-1$ or $\lambda_\mathcal{T}(G[S])$ or
$\lambda_\mathcal{T}(G[S])+1$. We distinguish the following cases:

\y {\bf \textit{Case~A.1}:} When the algorithm processes vertex
$v_n$, it uses $v_n$ to bridge two paths (operation bridge), that
is, $\lambda_\mathcal{T}(G)=\lambda_\mathcal{T}(G[S])-1$;
consequently, $\mathcal{P_{\mathcal{T}}}(G)$ is a minimum 1PC of
$G$, that is, $\lambda'_\mathcal{T}(G)=\lambda_\mathcal{T}(G)$.

\y \textit{Case~A.1.a}: Suppose that $\varepsilon'^{(n-1)}_i \leq
\varepsilon^{(n-1)}_i$, $d \leq i \leq n-1$. We show that the
algorithm computes a minimum 1PC $\mathcal{P_{\mathcal{T}}}(G)$ of
the graph $G$ having $\varepsilon^{(n)}_i$ endpoints $v_\kappa$
belonging to different paths with index $\kappa \in (i,n]$, $1
\leq i \leq n$, such that there is no other minimum 1PC
$\mathcal{P'_{\mathcal{T}}}(G)$ having $\varepsilon'^{(n)}_i$
endpoints $v_{\kappa'}$ belonging to different paths with index
$\kappa' \in (i,n]$ such that
$\varepsilon'^{(n)}_i>\varepsilon^{(n)}_i$, $1 \leq i \leq n$.
Clearly, vertex $v_n$ is an internal vertex of a path in any other
minimum 1PC $\mathcal{P'_{\mathcal{T}}}(G)$, otherwise removing it
from $\mathcal{P'_{\mathcal{T}}}(G)$ would result to a 1PC of
$G[S]$ of size $\leq \lambda_\mathcal{T}(G[S])-1$, a
contradiction. Assume that there exists a minimum 1PC
$\mathcal{P'_{\mathcal{T}}}(G)$ having an index, say, $k-1$, for
which we have $\varepsilon'^{(n)}_{k-1}$ endpoints $v_{\kappa'}$
belonging to different paths with index $\kappa' \in (k-1,n]$ such
that $\varepsilon'^{(n)}_{k-1}>\varepsilon^{(n)}_{k-1}$. Suppose
that $\varepsilon'^{(n)}_{k-1} - \varepsilon^{(n)}_{k-1}= 1$. Note
that $\varepsilon^{(n)}_1 \geq \varepsilon'^{(n)}_1$. Indeed, let
$\mathcal{P'_{\mathcal{T}}}(G)$ be a minimum 1PC of $G$ having
$\varepsilon'^{(n)}_1 =\varepsilon^{(n)}_1 +1$ endpoints
$v_{\kappa'}$ belonging to different paths with index $\kappa'
>1$. Suppose that vertex $v_1$ is an internal vertex in
$\mathcal{P_{\mathcal{T}}}(G)$. Then, the algorithm constructs
$\varepsilon^{(n)}_1$ paths while $\mathcal{P'_{\mathcal{T}}}(G)$
contains at least $\varepsilon^{(n)}_1+1$ paths, a contradiction.
Suppose that vertex $v_1$ is an endpoint in
$\mathcal{P_{\mathcal{T}}}(G)$. Note that, according to the
algorithm, if $v_1$ has degree greater or equal to one, then it
belongs to a path containing more than one vertex. Consequently,
the other endpoint of the path containing $v_1$ is one of the
$\varepsilon^{(n)}_1$ endpoints, and, thus, the algorithm
constructs $\varepsilon^{(n)}_1$ paths while
$\mathcal{P'_{\mathcal{T}}}(G)$ contains at least
$\varepsilon^{(n)}_1+1$ paths, a contradiction. Since
$\varepsilon^{(n)}_1 \geq \varepsilon'^{(n)}_1$ there exists a
vertex $v_j$, $1 \leq j < k-1$, such that $\varepsilon^{(n)}_j =
\varepsilon'^{(n)}_j$. This implies that vertex $v_{j+1}$ is the
right endpoint of a path $P$ in the minimum 1PC
$\mathcal{P_{\mathcal{T}}}(G)$ constructed by the algorithm.
Without loss of generality, we assume that $\varepsilon'^{(n)}_i =
\varepsilon^{(n)}_i + 1$, $j+1 \leq i < k-1$.

\y Let $P'=(\ldots, v_a, v_n, v_b, \ldots)$ be the path of
$\mathcal{P'_{\mathcal{T}}}(G)$ containing vertex $v_n$. Then,
$j+1<a$ and $j+1<b$. Indeed, suppose that at least one of $v_a$
and $v_b$ has index less or equal to $j+1$, say, $a<j+1$. Since
$v_av_n \in E(G)$ we have $v_{j+1}v_n \in E(G)$. Let $P=(\ldots,
v_c, v_n, v_d, \ldots)$ be the path of
$\mathcal{P_{\mathcal{T}}}(G)$ containing vertex $v_n$. Suppose
that $v_c<v_{j+1}$ and $v_d<v_{j+1}$. Then, due to the induction
hypothesis, both of the endpoints of $P$ have index greater than
$j+1$. However, according to the algorithm (operation bridge),
such an ordering of the endpoints cannot exist, a contradiction.
Suppose that $v_c>v_{j+1}$ and $v_d>v_{j+1}$. Then, due to the
induction hypothesis, at least one of the endpoints of $P$ should
have index greater than $j+1$. Again, according to the algorithm
(operation bridge), such an ordering of the endpoints cannot
exist, a contradiction. Suppose now that $v_c<v_{j+1}$ and
$v_d>v_{j+1}$. Then, due to the induction hypothesis, computing a
1PC of $G[S]$ we had case~(e) which is described in Section~3 and
$v_{j+1}=v_s$. However, in this case, vertex $v_{j+1}$ would not
be the right endpoint of a path in
$\mathcal{P_{\mathcal{T}}}(G[S])$, a contradiction.

\y Consequently, $j+1<a$ and $j+1<b$. Then, both endpoints of $P$
have indexes less than $j+1$. This implies that vertex $v_n$ has
bridged two non-terminal paths for which we have an endpoint of
one path between the endpoints of the other, which is a
contradiction according to Lemma~3.2.

\y Consequently, we have shown that there does not exist a minimum
1PC $\mathcal{P'_{\mathcal{T}}}(G)$ having an index, say, $k-1$,
for which we have $\varepsilon'^{(n)}_{k-1}$ endpoints
$v_{\kappa'}$ belonging to different paths with index $\kappa' \in
(k-1,n]$, where
$\varepsilon'^{(n)}_{k-1}>\varepsilon^{(n)}_{k-1}$.

\y Case~1.b: Suppose that $\varepsilon'^{(n-1)}_i =
\varepsilon^{(n-1)}_i +1$, $d \leq i < d'$ and
$\varepsilon'^{(n-1)}_i = \varepsilon^{(n-1)}_i$, $d' \leq i \leq
n-1$, where $d'$ is the index of the rightmost endpoint of a path
in $\mathcal{P'_{\mathcal{T}}}(G[S])$, and there exists a vertex
$v_q$ such that $\varepsilon^{(n-1)}_q > \varepsilon'^{(n-1)}_q$,
$1 \leq q < d$ and there exists no vertex $v_q'$, $1 \leq q' <d$,
such that $\varepsilon'^{(n-1)}_q>\varepsilon^{(n-1)}_q$. We show
that the algorithm computes a minimum 1PC
$\mathcal{P_{\mathcal{T}}}(G)$ of the graph $G$ having
$\varepsilon^{(n)}_i$ endpoints $v_\kappa$ belonging to different
paths with index $\kappa \in (i,n]$, $1 \leq i \leq n$, such that
there is no other minimum 1PC $\mathcal{P'_{\mathcal{T}}}(G)$
having $\varepsilon'^{(n)}_i$ endpoints $v_{\kappa'}$ belonging to
different paths with index $\kappa' \in (i,n]$, $1 \leq i \leq
\varrho$, such that $\varepsilon'^{(n)}_i>\varepsilon^{(n)}_i$, $1
\leq i < \varrho$, where $\varrho$ is the index of the rightmost
endpoint of a path in $\mathcal{P_{\mathcal{T}}}(G)$. Furthermore,
one of the following holds:
\begin{itemize}
\item[(i)] $\varepsilon'^{(n)}_i \leq \varepsilon^{(n)}_i$,
$\varrho \leq i \leq n$, or \item[(ii)] if $\varepsilon'^{(n)}_i =
\varepsilon^{(n)}_i +1$, $\varrho \leq i < \varrho'$ and
$\varepsilon'^{(n)}_i = \varepsilon^{(n)}_i$, $\varrho' \leq i
\leq n$, where $\varrho'$ is the index of the rightmost endpoint
of a path in $\mathcal{P'_{\mathcal{T}}}(G)$, then there exists a
vertex $v_z$ such that $\varepsilon^{(n)}_z >
\varepsilon'^{(n)}_z$, $1 \leq z < \varrho$ and there exists no
vertex $v_z'$, $1 \leq z' <\varrho$, such that
$\varepsilon'^{(n)}_z>\varepsilon^{(n)}_z$.

\end{itemize}

\y Suppose that there exists a minimum 1PC
$\mathcal{P'_{\mathcal{T}}}(G)$ such that
$\varepsilon^{(n)}_\varrho = \varepsilon'^{(n)}_\varrho =0$. Then,
similarly to Case~A.1.a, we show that there is no other minimum
1PC $\mathcal{P'_{\mathcal{T}}}(G)$ such that
$\varepsilon'^{(n)}_i>\varepsilon^{(n)}_i$, $1 \leq i < n$.

\y Suppose now that there exists a minimum 1PC
$\mathcal{P'_{\mathcal{T}}}(G)$ such that $\varepsilon'^{(n)}_i =
\varepsilon^{(n)}_i +1$, $\varrho \leq i < \varrho'$ and
$\varepsilon'^{(n)}_i = \varepsilon^{(n)}_i$, $\varrho' \leq i
\leq n$, where $\varrho'$ is the index of the rightmost endpoint
of a path in $\mathcal{P'_{\mathcal{T}}}(G)$. We show that there
exists a vertex $v_z$ such that $\varepsilon^{(n)}_z >
\varepsilon'^{(n)}_z$, $1 \leq z < \varrho$ and there exists no
vertex $v_{z'}$, $1 \leq z' <\varrho$, such that
$\varepsilon'^{(n)}_{z'}>\varepsilon^{(n)}_{z'}$.

\y Note that, there cannot exist a minimum 1PC
$\mathcal{P'_{\mathcal{T}}}(G)$ such that $\varepsilon'^{(n)}_i =
\varepsilon^{(n)}_i +2$, $\varrho \leq i < \varrho'$. Indeed,
suppose that there exists a minimum 1PC
$\mathcal{P'_{\mathcal{T}}}(G)$ such that $\varepsilon'^{(n)}_i =
\varepsilon^{(n)}_i +2$, $\varrho \leq i < \varrho'$. Consider the
case where $v_n$ belongs to a path in
$\mathcal{P'_{\mathcal{T}}}(G)$ and it is connected to vertices
$v_{a'}$ and $v_{b'}$ such that $\varrho<a'$ and $\varrho<b'$.
Then, removing $v_n$ from $\mathcal{P'_{\mathcal{T}}}(G)$ we
obtain a minimum 1PC of $G[S]$ such that
$\varepsilon'^{(n-1)}_\varrho=\varepsilon'^{(n)}_\varrho
+1=\varepsilon^{(n)}_\varrho+3$ or
$\varepsilon'^{(n-1)}_\varrho=\varepsilon'^{(n)}_\varrho
+2=\varepsilon^{(n)}_\varrho+4$. If $v_n$ belongs to a path in
$\mathcal{P_{\mathcal{T}}}(G)$ and it is connected to vertices
$v_a$ and $v_b$, then removing $v_n$ from
$\mathcal{P_{\mathcal{T}}}(G)$ we obtain a minimum 1PC of $G[S]$
such that $\varepsilon^{(n-1)}_\varrho \leq
\varepsilon^{(n)}_\varrho +2$. Thus, there exist three paths in
$\mathcal{P_{\mathcal{T}}}(G)$ such that there are no left
endpoints between their right endpoints in $\pi$, a contradiction.
Consider now the case where $v_n$ belongs to a path in
$\mathcal{P'_{\mathcal{T}}}(G)$ and it is connected to vertices
$v_{a'}$ and $v_{b'}$ such that $\varrho<a'$ and $\varrho>b'$.
Then, $v_nv_\varrho \in E(G)$. Note that $v_n$ belongs to a path
$P$ in $\mathcal{P_{\mathcal{T}}}(G)$ and it is connected to
vertices $v_a$ and $v_b$ such that $a<\varrho$ and $\varrho<b$.
Removing $v_n$ from $\mathcal{P_{\mathcal{T}}}(G)$ should result
to a minimum 1PC of $G[S]$ such that $\varepsilon^{(n-1)}_\varrho
= \varepsilon^{(n)}_\varrho + 1$. Consequently, both endpoints of
$P$ have index less than $\varrho$, which implies that there
exists a specific ordering of the endpoints of the paths in
$\mathcal{P_{\mathcal{T}}}(G[S])$, which, according to the
algorithm, is not possible, a contradiction. The case where $v_n$
belongs to a path in $\mathcal{P'_{\mathcal{T}}}(G)$ and it is
connected to vertices $v_{a'}$ and $v_{b'}$ such that $\varrho>a'$
and $\varrho>b'$ is similar.

\y Using similar arguments with Case~A.1.a, we show that there
exists no vertex $v_{z'}$, $1 \leq z' <\varrho$, such that
$\varepsilon'^{(n)}_{z'}>\varepsilon^{(n)}_{z'}$. Suppose that
$\varepsilon^{(n)}_i=\varepsilon'^{(n)}_i$, $1 \leq i <\varrho$.
Note that $v_n$ belongs to a path $P=(v_\ell, \ldots, v_a, v_n,
v_b, \ldots, v_r) \in \mathcal{P_{\mathcal{T}}}(G)$ such that
$a>\varrho$ and $b>\varrho$. Thus, $v_n$ belongs to a path
$P'=(v_{\ell'}, \ldots, v_{a'}, v_n, v_{b'}, \ldots, v_{r'}) \in
\mathcal{P'_{\mathcal{T}}}(G)$ such that $a'>\varrho$ and
$b'>\varrho$. If we remove $v_n$ from
$\mathcal{P'_{\mathcal{T}}}(G)$ then there exists a vertex
$v_z=v_{b'-1}$ such that
$\varepsilon'^{(n-1)}_z=\varepsilon'^{(n)}_z+1$. This implies that
$r'>b'$. Furthermore, if we remove $v_n$ from
$\mathcal{P_{\mathcal{T}}}(G)$ and we obtain
$\varepsilon^{(n-1)}_z=\varepsilon^{(n)}_z+1$, which implies that
$b=b'$, then $\varepsilon^{(n-1)}_i=\varepsilon'^{(n-1)}_i$, $1
\leq i <\varrho$, a contradiction. If we remove $v_n$ from
$\mathcal{P'_{\mathcal{T}}}(G)$ and we obtain
$\varepsilon^{(n-1)}_z=\varepsilon^{(n)}_z+2$, then there exists a
specific ordering of the endpoints of the paths in
$\mathcal{P_{\mathcal{T}}}(G[S])$ which, according to the
algorithm, is not possible, a contradiction. Consequently, there
exists a vertex $v_z$ such that $\varepsilon^{(n)}_z >
\varepsilon'^{(n)}_z$, $1 \leq z < \varrho$.

\y {\bf \textit{Case~A.2}:} When the algorithm processes vertex
$v_n$, it constructs a 1PC of size $\lambda_\mathcal{T}(G[S])-1$
of $\mathcal{P_{\mathcal{T}}}(G[S]-v_t)$ of $G[S]-v_t$, where
$v_t$ is the terminal vertex, and then connects the path $(v_t,
v_n)$ to an existing path. This operation is performed when vertex
$v_n$ sees the endpoints of at least one non-terminal path, say
$P_1=(v_r, \ldots, v_s)$, the terminal vertex $v_t$ and no other
endpoint of the terminal path, say, $P_2=(v_t, \ldots, v_\ell)$.
Then, the terminal path, $P_2$, has the same endpoints as in
$\mathcal{P_{\mathcal{T}}}(G[S])$, the vertices of $P_1$ become
internal vertices of $P_2$, while all the other paths remain the
same. Note that when the connect operation is performed, it may
use a vertex of the terminal path in order to increase the value
of an $\varepsilon^{(n-1)}_i$, $1 \leq i < n-1$, and, in this
case, $\varepsilon^{(n-1)}_d<\varepsilon'^{(n-1)}_d$. Then, for
the endpoints of the terminal path, say, $v_t \in \mathcal{T}$ and
$v_\ell$, we have $v_t<v_\ell$. Consequently, since vertex $v_n$
sees the endpoints of $P_1$, the terminal vertex $v_t$ and no
other endpoint of the terminal path $P_2$, if operation connect
was called previously, it cannot have used a vertex of the
terminal path and, thus,
$\varepsilon^{(n)}_d<\varepsilon'^{(n)}_d$ cannot hold.
Consequently, $\varepsilon'^{(n-1)}_i \leq \varepsilon^{(n-1)}_i$,
$d \leq i \leq n-1$.

\y The above procedure results to a 1PC of $G$ of size
$\lambda_\mathcal{T}(G)=\lambda_\mathcal{T}(G[S])-1$;
consequently, $\mathcal{P_{\mathcal{T}}}(G)$ is a minimum 1PC of
$G$, that is, $\lambda'_\mathcal{T}(G)=\lambda_\mathcal{T}(G)$.

\y We show that the algorithm computes a minimum 1PC
$\mathcal{P_{\mathcal{T}}}(G)$ of the graph $G$ having
$\varepsilon^{(n)}_i$ endpoints $v_\kappa$ belonging to different
paths with index $\kappa \in (i,n]$, $1 \leq i \leq n$, such that
there is no other minimum 1PC $\mathcal{P'_{\mathcal{T}}}(G)$
having $\varepsilon'^{(n)}_i$ endpoints $v_{\kappa'}$ belonging to
different paths with index $\kappa' \in (i,n]$ such that
$\varepsilon'^{(n)}_i>\varepsilon^{(n)}_i$, $1 \leq i \leq n$.
Clearly, vertex $v_n$ is an internal vertex of a path in any other
minimum 1PC $\mathcal{P'_{\mathcal{T}}}(G)$, otherwise removing it
from $\mathcal{P'_{\mathcal{T}}}(G)$ would result to a 1PC of
$G[S]$ of size less or equal to $\lambda_\mathcal{T}(G[S])-1$, a
contradiction. Suppose that there exists a minimum 1PC
$\mathcal{P'_{\mathcal{T}}}(G)$ having an index, say, $k-1$, such
that $\varepsilon'^{(n)}_{k-1}>\varepsilon^{(n)}_{k-1}$ and $1
\leq k-1 < t-1$. This implies that
$\varepsilon'^{(n)}_k=\varepsilon^{(n)}_k$ and there exists a
vertex $v_{k'-1}$ such that $k'-1<k-1$ and
$\varepsilon'^{(n)}_{k'-1}=\varepsilon^{(n)}_{k'-1}$. Clearly,
$v_{k'}v_n \notin E(G)$. Removing $v_n$ from
$\mathcal{P'_{\mathcal{T}}}(G)$ results to a minimum 1PC of $G[S]$
having at least two free neighbors of $v_n$ as endpoints belonging
to different paths, say, $v_f$ and $v_g$; suppose that at least
one of them has index greater than $k$. Then,
$\varepsilon'^{(n-1)}_{k-1}>\varepsilon^{(n-1)}_{k-1}$, a
contradiction. Thus, $v_{k'}<v_f<v_k$ and $v_{k'}<v_g<v_k$ and the
right endpoint of the path that they belong, has also index less
than $k$. Thus,
$\varepsilon'^{(n-1)}_{k'}=\varepsilon'^{(n)}_{k'}+1=\varepsilon^{(n)}_{k'}+1+1=\varepsilon^{(n-1)}_{k'}+1$
or
$\varepsilon'^{(n-1)}_{k'}=\varepsilon'^{(n)}_{k'}+2=\varepsilon^{(n)}_{k'}+1+2=\varepsilon^{(n-1)}_{k'}+2$,
a contradiction. Suppose now that $t \leq k-1 \leq n$. Since there
cannot exist a vertex $v_\ell$ such that $1 \leq \ell < t-1$ and
$\varepsilon'^{(n)}_\ell>\varepsilon^{(n)}_\ell$, there exists a
vertex $v_{k'-1}$ such that $k'-1<k-1$ and
$\varepsilon'^{(n)}_{k'-1}=\varepsilon^{(n)}_{k'-1}$ and $t \leq
k'$. Clearly, $v_{k'}v_n \in E(G)$. If $v_s$ is the right endpoint
of $P_1$, then $k-1<s$ and thus $k'-1<s$. However, according to
the algorithm, there cannot exist an endpoint between vertices
$v_t$ and $v_s$, a contradiction.

\y {\bf \textit{Case~A.3}:} When the algorithm processes vertex
$v_n$, it constructs a 1PC of size $\lambda_\mathcal{T}(G[S])$ of
$\mathcal{P_{\mathcal{T}}}(G[S]-v_t)$ of $G[S]-v_t$, where $v_t$
is the terminal vertex, it connects $v_t$ to the leftmost left
endpoint it sees, and it uses $v_n$ to bridge two paths. This
operation is performed when vertex $v_n$ sees the endpoints of at
least one non-terminal path, say $P_1=(v_r, \ldots, v_s)$, the
terminal vertex $v_t$ of the terminal path, say, $P_2=(v_t, v_j,
\ldots, v_\ell)$ and an internal vertex $v_j$ of $P_2$, and it
does not see $v_\ell$. Then, the terminal path, $P_2$, has the
same endpoints as in $\mathcal{P_{\mathcal{T}}}(G[S])$, the
vertices of $P_1$ become internal vertices of $P_2$, while all the
other paths remain the same. Recall that, when the connect
operation is performed, it may use a vertex of the terminal path
in order to increase the value of an $\varepsilon^{(n-1)}_i$, $1
\leq i < n-1$, and, in this case,
$\varepsilon^{(n-1)}_d<\varepsilon'^{(n-1)}_d$. Then, for the
endpoints of the terminal path, say, $v_t \in \mathcal{T}$ and
$v_\ell$, we have $v_t<v_\ell$. Consequently, since vertex $v_n$
sees the endpoints of $P_1$, the terminal vertex $v_t$ and not
$v_\ell$, if operation connect was called previously, it cannot
have used a vertex of the terminal path and, thus,
$\varepsilon^{(n-1)}_d<\varepsilon'^{(n-1)}_d$ cannot hold.
Consequently, $\varepsilon'^{(n-1)}_i \leq \varepsilon^{(n-1)}_i$,
$d \leq i \leq n-1$.

\y Consider the case where vertex $v_n$ sees the endpoints of only
one non-terminal path, that is, of $P_1$. If applying the
algorithm to $G[S]-v_t$ results to a 1PC of size
$\lambda_\mathcal{T}(G[S])-1$ then it contains a path with one
endpoint $v_k$ such that $t<k$. Indeed, suppose that there does
not exist such a path and $P_1$ consists of more than one vertex.
This implies that all vertices of $P_1$ have bridged two paths and
therefore we obtain a 1PC of size less than
$\lambda_\mathcal{T}(G[S])-1$, a contradiction. Suppose now that
the algorithm does not construct a path in the 1PC of $G[S]-v_t$
with one endpoint $v_k$ such that $t<k$ and let the path $P_1$
consist of one vertex, say, $v_{k'}$. This implies that vertex
$v_{k'}$ has bridged two paths and the same holds for every vertex
$v_i$, $t \leq i \leq n-1$. Thus, if $v_{k'}$ is removed from
$G[S]-v_t$, the algorithm would construct a minimum 1PC of size
$\lambda_\mathcal{T}(G[S])$. Since the size of
$\mathcal{P_{\mathcal{T}}}(G[S])$ constructed by the algorithm is
$\lambda_\mathcal{T}(G[S])$, removing $v_t$ and $v_{k'}$ results
to a 1PC of size $\lambda_\mathcal{T}(G[S])-1$. Consequently,
$v_{k'}$ cannot be used for bridging two paths in
$\mathcal{P_{\mathcal{T}}}(G[S]-v_t)$. This implies that the 1PC
of $G[S]-v_t$ constructed by the algorithm contains a path with an
endpoint $v_k$ such that $t<k$. It is easy to see that if vertex
$v_n$ sees the endpoints of more than one non-terminal path,
applying the algorithm to $G[S]-v_t$ results to a 1PC of size
$\lambda_\mathcal{T}(G[S])-1$ having a path with one endpoint
$v_k$ such that $t<k$.

\y The above procedure results to a 1PC of $G$ of size
$\lambda_\mathcal{T}(G)=\lambda_\mathcal{T}(G[S])-1$;
consequently, $\mathcal{P_{\mathcal{T}}}(G)$ is a minimum 1PC of
$G$, that is, $\lambda'_\mathcal{T}(G)=\lambda_\mathcal{T}(G)$.

\y Using similar arguments as in Case~A.2, we show that the
algorithm computes a minimum 1PC $\mathcal{P_{\mathcal{T}}}(G)$ of
the graph $G$ having $\varepsilon^{(n)}_i$ endpoints $v_\kappa$
belonging to different paths with index $\kappa \in (i,n]$, $1
\leq i \leq n$, such that there is no other minimum 1PC
$\mathcal{P'_{\mathcal{T}}}(G)$ having $\varepsilon'^{(n)}_i$
endpoints $v_{\kappa'}$ belonging to different paths with index
$\kappa' \in (i,n]$ such that
$\varepsilon'^{(n)}_i>\varepsilon^{(n)}_i$, $1 \leq i \leq n$.

\y {\bf \textit{Case~A.4}:} When the algorithm processes vertex
$v_n$, it connects $v_n$ to a path, that is,
$\lambda_\mathcal{T}(G)=\lambda_\mathcal{T}(G[S])$. Suppose that
there exists a 1PC $\mathcal{P'_{\mathcal{T}}}(G)$ of size
$\lambda_\mathcal{T}(G[S])-1$, that is, vertex $v_n$ is an
internal vertex of a path $P$ in $\mathcal{P'_{\mathcal{T}}}(G)$.
We distinguish the following cases:

(i) $P=(v_k, \ldots, v_r, v_n, v_s, \ldots, v_\ell)$. Removing
$v_n$ from $P$ results to a minimum 1PC of $G[S]$ having two
(free) neighbors of $v_n$ as endpoints belonging to different
paths. Since the algorithm does not use $v_n$ to bridge two paths,
the constructed minimum 1PC of $G[S]$ does not have two (free)
neighbors of $v_n$ as endpoints belonging to different paths.
Consequently, there is a minimum 1PC of $G[S]$ for which there
exists an index $i$ such that
$\varepsilon'^{(n)}_i>\varepsilon^{(n)}_i$, a contradiction.

(ii) $P=(v_t, v_n, \ldots, v_b)$, where $v_t$ is the terminal
vertex.  Removing $v_n$ and $v_t$ from $P$ results to a minimum
1PC of $G[S]$ of size $\lambda_\mathcal{T}(G[S])-1$, a
contradiction. Indeed, since the algorithm does not use $v_n$ to
bridge two paths, removing $v_t$ from $G[S]$ results to
$\lambda_\mathcal{T}(G[S])$ paths.

\y Consequently, there does not exist a 1PC
$\mathcal{P'_{\mathcal{T}}}(G)$ of size
$\lambda_\mathcal{T}(G[S])-1$, and, thus, the 1PC constructed by
the algorithm is minimum.

\y \textit{Case~A.4.a}: Suppose that $\varepsilon'^{(n-1)}_i \leq
\varepsilon^{(n-1)}_i$, $d \leq i \leq n-1$. We show that the
algorithm computes a minimum 1PC $\mathcal{P_{\mathcal{T}}}(G)$ of
the graph $G$ having $\varepsilon^{(n)}_i$ endpoints $v_\kappa$
belonging to different paths with index $\kappa \in (i,n]$, $1
\leq i \leq n$, such that there is no other minimum 1PC
$\mathcal{P'_{\mathcal{T}}}(G)$ having $\varepsilon'^{(n)}_i$
endpoints $v_{\kappa'}$ belonging to different paths with index
$\kappa' \in (i,n]$ such that
$\varepsilon'^{(n)}_i>\varepsilon^{(n)}_i$, $1 \leq i < \varrho$.
Furthermore, one of the following holds:
\begin{itemize}
\item[(i)] $\varepsilon'^{(n)}_i \leq \varepsilon^{(n)}_i$,
$\varrho \leq i \leq n$, or \item[(ii)] if $\varepsilon'^{(n)}_i =
\varepsilon^{(n)}_i +1$, $\varrho \leq i < \varrho'$ and
$\varepsilon'^{(n)}_i = \varepsilon^{(n)}_i$, $\varrho' \leq i
\leq n$ then there exists a vertex $v_z$ such that
$\varepsilon^{(n)}_z > \varepsilon'^{(n)}_z$, $1 \leq z < \varrho$
and there exists no vertex $v_{z'}$, $1 \leq z' <\varrho$, such
that $\varepsilon'^{(n)}_{z'}>\varepsilon^{(n)}_{z'}$.

\end{itemize}

\y Suppose that $v_n$ is not an endpoint in
$\mathcal{P_{\mathcal{T}}}(G)$; let $P=(\ldots, v_{a}, v_n, v_{b},
\ldots)$. According to operation connect, we break the terminal
path of $\mathcal{P_{\mathcal{T}}}(G[S])$, which has the terminal
vertex, $v_t$, as its right endpoint. Note that the terminal
vertex is the second rightmost endpoint in
$\mathcal{P_{\mathcal{T}}}(G[S])$ (see Section~3). The second
rightmost endpoint of $\mathcal{P_{\mathcal{T}}}(G)$ has index
greater than $t$, and $v_t$ becomes a left endpoint of a path in
$\mathcal{P_{\mathcal{T}}}(G)$. Assume that there exists a minimum
1PC $\mathcal{P'_{\mathcal{T}}}(G)$ having an index, say, $k-1$,
for which we have $\varepsilon'^{(n)}_{k-1}$ endpoints
$v_{\kappa'}$ belonging to different paths with index $\kappa' \in
(k-1,n]$, where
$\varepsilon'^{(n)}_{k-1}>\varepsilon^{(n)}_{k-1}$. Similarly, to
Case~A.1, there exists a vertex $v_j$, $1 \leq j < k-1$, such that
$\varepsilon^{(n)}_j = \varepsilon'^{(n)}_j$.

\y Suppose that $v_n$ is an endpoint of a path $P'=(v_n, v_{a'},
\ldots) \in \mathcal{P'_{\mathcal{T}}}(G)$. Then, since
$\varrho'=n$, $\varepsilon'^{(n)}_i = \varepsilon^{(n)}_i +1$,
$\varrho \leq i \leq n$. Note that, there cannot exist a minimum
1PC $\mathcal{P'_{\mathcal{T}}}(G)$ such that
$\varepsilon'^{(n)}_i = \varepsilon^{(n)}_i +2$, $\varrho \leq i
\leq n$. Indeed, suppose that there exists a minimum 1PC
$\mathcal{P'_{\mathcal{T}}}(G)$ such that $\varepsilon'^{(n)}_i =
\varepsilon^{(n)}_i +2$, $\varrho \leq i \leq n$. If we remove
$v_n$ from $\mathcal{P'_{\mathcal{T}}}(G)$ we obtain
$\varepsilon'^{(n-1)}_\varrho=\varepsilon'^{(n)}_\varrho$ or
$\varepsilon'^{(n-1)}_\varrho=\varepsilon'^{(n)}_\varrho-1$.
However, $\varepsilon'^{(n)}_\varrho=\varepsilon^{(n)}_\varrho+2$
and $\varepsilon^{(n)}_\varrho=\varepsilon^{(n-1)}_\varrho=0$, a
contradiction. According to the connect operation, $v_nv_{j+1}
\notin E(G)$, thus $v_{a'}>v_{j+1}$. If we remove $v_n$ from
$\mathcal{P'_{\mathcal{T}}}(G)$ we obtain
$\varepsilon'^{(n-1)}_{j+1}=\varepsilon'^{(n)}_{j+1}=\varepsilon^{(n)}_{j+1}+1$.
However, $\varepsilon^{(n)}_{j+1}=\varepsilon^{(n-1)}_{j+1}$ or
$\varepsilon^{(n)}_{j+1}=\varepsilon^{(n-1)}_{j+1}+1$, a
contradiction. Consequently, there exists no vertex $v_{z'}$, $1
\leq z' <\varrho$, such that
$\varepsilon'^{(n)}_{z'}>\varepsilon^{(n)}_{z'}$. Suppose that
$\varepsilon'^{(n)}_i=\varepsilon^{(n)}_i$, $1 \leq i <\varrho$.
Let $v_r$ be the new endpoint created by the connect operation.
Again, since $v_nv_{r-1} \notin E(G)$, $v_{a'}>v_{r-1}$ and if we
remove $v_n$ from $\mathcal{P'_{\mathcal{T}}}(G)$ we obtain
$\varepsilon'^{(n-1)}_{r-1}=\varepsilon'^{(n)}_{r-1}=\varepsilon^{(n)}_{r-1}$.
However, $\varepsilon^{(n)}_{r-1}=\varepsilon^{(n-1)}_{r-1}+1$, a
contradiction. Consequently, if $\varepsilon'^{(n)}_i =
\varepsilon^{(n)}_i +1$, $\varrho \leq i < \varrho'$ and
$\varepsilon'^{(n)}_i = \varepsilon^{(n)}_i$, $\varrho' \leq i
\leq n$ then there exists a vertex $v_z$ such that
$\varepsilon^{(n)}_z > \varepsilon'^{(n)}_z$, $1 \leq z < \varrho$
and there exists no vertex $v_{z'}$, $1 \leq z' <\varrho$, such
that $\varepsilon'^{(n)}_{z'}>\varepsilon^{(n)}_{z'}$. Suppose
that the second rightmost endpoint in
$\mathcal{P_{\mathcal{T}}}(G)$, say, $v_f$, has index less than
the second rightmost endpoint in $\mathcal{P'_{\mathcal{T}}}(G)$,
say, $v_{f'}$, that is, $v_f<v_{f'}$. Then,
$\varepsilon^{(n-1)}_{f-1}=\varepsilon^{(n)}_{f-1}-1=\varepsilon'^{(n)}_{f-1}-2$
and $\varepsilon'^{(n-1)}_{f-1}=\varepsilon'^{(n)}_{f-1}-1$, a
contradiction.

\y Suppose that $v_n$ is not an endpoint in
$\mathcal{P'_{\mathcal{T}}}(G)$; let $P'=(\ldots, v_{a'}, v_n,
v_{b'}, \ldots)$. Clearly, one of $v_{a'}, v_{b'}$ is a vertex
that could not be an endpoint in $\mathcal{P'_{\mathcal{T}}}(G)$.
We show that $\varrho' \leq \varrho$. Suppose that $\varrho'
> \varrho$. Since $\varepsilon^{(n-1)}_\varrho=\varepsilon^{(n)}_\varrho=\varepsilon'^{(n)}_\varrho-1=0$,
then we have
$\varepsilon'^{(n-1)}_\varrho=\varepsilon'^{(n)}_\varrho-1$, which
implies that the new endpoint in $\mathcal{P'_{\mathcal{T}}}(G)$
has index greater than the new endpoint created in
$\mathcal{P_{\mathcal{T}}}(G)$, a contradiction. It is easy to see
that there cannot exist a a minimum 1PC
$\mathcal{P'_{\mathcal{T}}}(G)$ having an index, say, $k-1$, for
which we have $\varepsilon'^{(n)}_{k-1}$ endpoints $v_{\kappa'}$
belonging to different paths with index $\kappa' \in (k-1,n]$ such
that $\varepsilon'^{(n)}_{k-1}=\varepsilon'^{(n-1)}_{k-1}+2$. Let
$v_t$ be the terminal vertex. We have
$\varepsilon^{(n)}_{t-1}=\varepsilon'^{(n-1)}_{t-1}$. It is easy
to see that there cannot exist a minimum 1PC
$\mathcal{P'_{\mathcal{T}}}(G)$ having an index, say, $k-1$, $k-1
\leq t-1$, for which we have $\varepsilon'^{(n)}_{k-1}$ endpoints
$v_{\kappa'}$ belonging to different paths with index $\kappa' \in
(k-1,n]$, such that
$\varepsilon'^{(n)}_{k-1}>\varepsilon^{(n)}_{k-1}$.

\y Suppose that $v_n$ is the right endpoint of a path $P=(v_n,
v_a, \ldots) \in \mathcal{P_{\mathcal{T}}}(G)$. Then, there cannot
exist a minimum 1PC $\mathcal{P'_{\mathcal{T}}}(G)$ such that $1=
\varepsilon^{(n)}_{\varrho-1} < \varepsilon'^{(n)}_{\varrho-1}$.
We show that the algorithm computes a minimum 1PC
$\mathcal{P_{\mathcal{T}}}(G)$ of the graph $G$ having
$\varepsilon^{(n)}_i$ endpoints $v_\kappa$ belonging to different
paths with index $\kappa \in (i,n]$, $1 \leq i \leq n$, such that
there is no other minimum 1PC $\mathcal{P'_{\mathcal{T}}}(G)$
having $\varepsilon'^{(n)}_i$ endpoints $v_{\kappa'}$ belonging to
different paths with index $\kappa' \in (i,n]$ such that
$\varepsilon'^{(n)}_i>\varepsilon^{(n)}_i$, $1 \leq i \leq n$.
Assume that there exists a minimum 1PC
$\mathcal{P'_{\mathcal{T}}}(G)$ having an index, say, $k-1$, for
which we have $\varepsilon'^{(n)}_{k-1}$ endpoints $v_{\kappa'}$
belonging to different paths with index $\kappa' \in (k-1,n]$,
where $\varepsilon'^{(n)}_{k-1}>\varepsilon^{(n)}_{k-1}$.
Similarly, to Case~1, there exists a vertex $v_j$, $1 \leq j <
k-1$, such that $\varepsilon^{(n)}_j = \varepsilon'^{(n)}_j$.

\y Suppose that $v_n$ is an endpoint of a path $P'=(v_n, v_{a'},
\ldots) \in \mathcal{P'_{\mathcal{T}}}(G)$. Note that if
$v_{a'}<v_{j+1}$, then $v_{j+1}v_n \in E(G)$ and $v_k$ should be
the terminal vertex belonging to a non-trivial path, otherwise
$\mathcal{P_{\mathcal{T}}}(G)$ would not be minimum. Thus, if
$v_{j+1}$ is not the terminal vertex, $v_a<v_{j+1}$ or
$v_a=v_{j+1}$. Then, if we remove $v_n$ from
$\mathcal{P'_{\mathcal{T}}}(G)$ and
$\mathcal{P_{\mathcal{T}}}(G)$, we obtain
$\varepsilon^{(n-1)}_{j+1}=\varepsilon^{(n)}_{j+1}-1=\varepsilon'^{(n)}_{j+1}-2$
and $\varepsilon'^{(n-1)}_{j+1}=\varepsilon'^{(n)}_{j+1}-1$; thus,
$\varepsilon^{(n-1)}_{j+1}=\varepsilon'^{(n-1)}_{j+1}-1$, a
contradiction. On the other hand, if $v_{a'}>v_{j+1}$, then, if we
remove $v_n$ from $\mathcal{P_{\mathcal{T}}}(G)$ and
$\mathcal{P'_{\mathcal{T}}}(G)$, we obtain
$\varepsilon^{(n-1)}_{j+1}=\varepsilon^{(n)}_{j+1}-1=\varepsilon'^{(n)}_{j+1}-2$
or
$\varepsilon^{(n-1)}_{j+1}=\varepsilon^{(n)}_{j+1}=\varepsilon'^{(n)}_{j+1}-1$.
Also, $\varepsilon'^{(n-1)}_{j+1}=\varepsilon'^{(n)}_{j+1}$; thus,
$\varepsilon^{(n-1)}_{j+1}=\varepsilon'^{(n-1)}_{j+1}-1$, a
contradiction.

\y Suppose that $v_n$ is not an endpoint in
$\mathcal{P'_{\mathcal{T}}}(G)$; let $P'=(\ldots, v_{a'}, v_n,
v_{b'}, \ldots)$. If $v_{a'}v_{b'} \in E(G)$ then if we remove
$v_n$ from $\mathcal{P'_{\mathcal{T}}}(G)$ and
$\mathcal{P'_{\mathcal{T}}}(G)$, we obtain
$\varepsilon^{(n-1)}_{j+1}=\varepsilon^{(n)}_{j+1}-1=\varepsilon'^{(n)}_{j+1}-2$
or
$\varepsilon^{(n-1)}_{j+1}=\varepsilon^{(n)}_{j+1}=\varepsilon'^{(n)}_{j+1}-1$.
Also, $\varepsilon'^{(n-1)}_{j+1}=\varepsilon'^{(n)}_{j+1}$; thus,
$\varepsilon^{(n-1)}_{j+1}=\varepsilon'^{(n-1)}_{j+1}-1$, a
contradiction. Consequently, $v_{a'}v_{b'} \notin E(G)$; however,
we have shown that $v_n$ becomes an endpoint in
$\mathcal{P_{\mathcal{T}}}(G)$ only when a new right endpoint
cannot be created by making $v_n$ an internal vertex, a
contradiction.

\y \textit{Case~A.4.b}: Suppose that $\varepsilon'^{(n-1)}_i =
\varepsilon^{(n-1)}_i +1$, $d \leq i < d'$ and
$\varepsilon'^{(n-1)}_i = \varepsilon^{(n-1)}_i$, $d' \leq i \leq
n-1$ and there exists a vertex $v_q$ such that
$\varepsilon^{(n-1)}_q > \varepsilon'^{(n-1)}_q$, $1 \leq q < d$
and there exists no vertex $v_{q'}$, $1 \leq q' <d$, such that
$\varepsilon'^{(n-1)}_{q'}>\varepsilon^{(n-1)}_{q'}$. We show that
the algorithm computes a minimum 1PC
$\mathcal{P_{\mathcal{T}}}(G)$ of the graph $G$ having
$\varepsilon^{(n)}_i$ endpoints $v_\kappa$ belonging to different
paths with index $\kappa \in (i,n]$, $1 \leq i \leq n$, such that
there is no other minimum 1PC $\mathcal{P'_{\mathcal{T}}}(G)$
having $\varepsilon'^{(n)}_i$ endpoints $v_{\kappa'}$ belonging to
different paths with index $\kappa' \in (i,n]$ such that
$\varepsilon'^{(n)}_i>\varepsilon^{(n)}_i$, $1 \leq i \leq n$.

\y Assume that there exists a minimum 1PC
$\mathcal{P'_{\mathcal{T}}}(G)$ having an index, say, $k-1$, for
which we have $\varepsilon'^{(n)}_{k-1}$ endpoints $v_{\kappa'}$
belonging to different paths with index $\kappa' \in (k-1,n]$,
where $\varepsilon'^{(n)}_{k-1}>\varepsilon^{(n)}_{k-1}$.
Similarly, to Case~A.1, there exists a vertex $v_j$, $1 \leq j <
k-1$, such that $\varepsilon^{(n)}_j = \varepsilon'^{(n)}_j$.
Using similar arguments as in Case~A.4.a, we show that $v_n$ is an
endpoint of a path $P'=(v_n, v_{a'}, \ldots) \in
\mathcal{P'_{\mathcal{T}}}(G)$ and that there cannot exist a
minimum 1PC $\mathcal{P'_{\mathcal{T}}}(G)$ having an index, say,
$k-1$, for which we have $\varepsilon'^{(n)}_{k-1}$ endpoints
$v_{\kappa'}$ belonging to different paths with index $\kappa' \in
(k-1,n]$, where
$\varepsilon'^{(n)}_{k-1}>\varepsilon^{(n)}_{k-1}$.

\y {\bf \textit{Case~A.5}:} When the algorithm processes vertex
$v_n$, it inserts $v_n$ into a path, that is,
$\lambda_\mathcal{T}(G)=\lambda_\mathcal{T}(G[S])$. This implies
that $\forall i \geq d$ we have $\varepsilon^{(n-1)}_i \leq 1$.
Suppose that there exists a 1PC $\mathcal{P'_{\mathcal{T}}}(G)$ of
size $\lambda_\mathcal{T}(G[S])-1$, that is, vertex $v_n$ is an
internal vertex of a path $P$ in $\mathcal{P'_{\mathcal{T}}}(G)$.
Then, removing vertex $v_n$ from $\mathcal{P'_{\mathcal{T}}}(G)$
results to a 1PC of $G[S]$ of size $\lambda_\mathcal{T}(G[S])$,
and, thus, minimum, such that there exists an index $i$, $i \geq
d$, for which $\varepsilon'^{(n-1)}_i=2$, a contradiction.

\y Consequently, there does not exist a 1PC
$\mathcal{P'_{\mathcal{T}}}(G)$ of size
$\lambda_\mathcal{T}(G[S])-1$, and, thus, the 1PC constructed by
the algorithm is minimum.

\y \textit{Case~A.5.a}: Suppose that $\varepsilon'^{(n-1)}_i \leq
\varepsilon^{(n-1)}_i$, $d \leq i \leq n-1$. We show that the
algorithm computes a minimum 1PC $\mathcal{P_{\mathcal{T}}}(G)$ of
the graph $G$ having $\varepsilon^{(n)}_i$ endpoints $v_\kappa$
belonging to different paths with index $\kappa \in (i,n]$, $1
\leq i \leq n$, such that there is no other minimum 1PC
$\mathcal{P'_{\mathcal{T}}}(G)$ having $\varepsilon'^{(n)}_i$
endpoints $v_{\kappa'}$ belonging to different paths with index
$\kappa' \in (i,n]$ such that
$\varepsilon'^{(n)}_i>\varepsilon^{(n)}_i$, $1 \leq i \leq n$.

\y Assume that there exists a minimum 1PC
$\mathcal{P'_{\mathcal{T}}}(G)$ having an index, say, $k-1$, for
which we have $\varepsilon'^{(n)}_{k-1}$ endpoints $v_{\kappa'}$
belonging to different paths with index $\kappa' \in (k-1,n]$,
where $\varepsilon'^{(n)}_{k-1}>\varepsilon^{(n)}_{k-1}$.
Similarly, to Case~A.1, there exists a vertex $v_j$, $1 \leq j <
k-1$, such that $\varepsilon^{(n)}_j = \varepsilon'^{(n)}_j$.

\y Suppose that $v_n$ is an endpoint of a path $P'=(v_n, v_t) \in
\mathcal{P'_{\mathcal{T}}}(G)$, such that $v_t \in \mathcal{T}$.
Then, $v_tv_n \in E(G)$ and the size of a minimum 1PC of
$G[S]-v_t$ is $\lambda_\mathcal{T}(G)-1$, a contradiction.

\y Suppose that $v_n$ is an endpoint of a path $P'=(v_n, v_{a'},
\ldots, v_{b'}) \in \mathcal{P'_{\mathcal{T}}}(G)$, such that
$v_{a'} \notin \mathcal{T}$. Then, removing vertex $v_n$ from
$\mathcal{P'_{\mathcal{T}}}(G)$ results to a 1PC of $G[S]$ of size
$\lambda_\mathcal{T}(G[S])$, and, thus, minimum, such that there
exists an index $i$ for which $\varepsilon'^{(n-1)}_i=1$, $i \geq
d$. Then, $d=d'$, which is equal to the index of the terminal
vertex, and $P'$ is the terminal path such that its left endpoint
in $\mathcal{P'_{\mathcal{T}}}(G[S])$, that is, vertex $v_{a'}$,
has index greater than the index of the left endpoint of the
terminal path in $\mathcal{P_{\mathcal{T}}}(G[S])$. Note that,
$v_{a'}$ cannot be an endpoint in
$\mathcal{P'_{\mathcal{T}}}(G[S])$ since $\varepsilon'^{(n-1)}_i
\leq \varepsilon^{(n-1)}_i$, $d \leq i \leq n-1$, and, thus, a 1PC
of $G[S]$ having $v_{a'}$ as an endpoint cannot be minimum.
However, removing $v_n$ from $\mathcal{P'_{\mathcal{T}}}(G)$
results to a 1PC of $G[S]$ of size $\lambda_\mathcal{T}(G[S])$
having $v_{a'}$ as an endpoint, a contradiction.

\y Suppose now that $v_n$ is not an endpoint in
$\mathcal{P'_{\mathcal{T}}}(G)$; let $P'=(\ldots, v_{a'}, v_n,
v_{b'}, \ldots)$. If $v_{a'}v_{b'} \in E(G)$ then if we remove
$v_n$ from $\mathcal{P'_{\mathcal{T}}}(G)$ and
$\mathcal{P'_{\mathcal{T}}}(G)$, we obtain
$\varepsilon^{(n-1)}_{k-1}=\varepsilon^{(n)}_{k-1}=\varepsilon'^{(n)}_{k-1}-1$
and $\varepsilon'^{(n-1)}_{k-1}=\varepsilon'^{(n)}_{k-1}$, a
contradiction. Consequently, $v_{a'}v_{b'} \notin E(G)$. Suppose
that the value of d-connectivity of
$\mathcal{P_{\mathcal{T}}}(G[S])$ is $c$; then the value of
d-connectivity of $\mathcal{P_{\mathcal{T}}}(G)$ is $c+2$.
However, the corresponding value of
$\mathcal{P'_{\mathcal{T}}}(G)$ is not increased by vertices
$v_{a'}, v_n$ and $v_{b'}$, since $v_{a'}$ and $v_{b'}$ are
internal vertices not successive into a path in a 1PC of $G[S]$
and there exist two vertices connected to $v_{a'}$ and $v_{b'}$ in
$\mathcal{P_{\mathcal{T}}}(G[S])$, say, $v_f$ and $v_g$,
respectively, for which $d(v_f)$ and $d(v_g)$ are reduced, and,
thus, they reduce the d-connectivity by two. In order to obtain
$c+2$ for $\mathcal{P'_{\mathcal{T}}}(G)$ the vertices of
$V(G)-\{v_{a'}, v_{b'}, v_n\}$ must increase the d-connectivity by
two. However, the size of $\mathcal{P'_{\mathcal{T}}}(G)$ is also
$\lambda_\mathcal{T}(G)$ and vertices $v_{a'}, v_{b'}$ and $v_n$
are also internal in $\mathcal{P_{\mathcal{T}}}(G)$. Thus,
increasing the d-connectivity by two is not possible and we have a
contradiction. Note that $v_fv_g \notin E(G)$; otherwise we would
have also $v_nv_f \in E(G)$ and $v_nv_g \in E(G)$ and there would
exist a 1PC having the same endpoints as
$\mathcal{P'_{\mathcal{T}}}(G)$ and containing a path $P=(\ldots,
v_f, v_n, v_g, \ldots)$ with $v_fv_g \in E(G)$, a contradiction.

\y \textit{Case~A.5.b}: Suppose that $\varepsilon'^{(n-1)}_i =
\varepsilon^{(n-1)}_i +1$, $d \leq i < d'$ and
$\varepsilon'^{(n-1)}_i = \varepsilon^{(n-1)}_i$, $d' \leq i \leq
n-1$ and there exists a vertex $v_q$ such that
$\varepsilon^{(n-1)}_q > \varepsilon'^{(n-1)}_q$, $1 \leq q < d$
and there exists no vertex $v_{q'}$, $1 \leq q' <d$, such that
$\varepsilon'^{(n-1)}_{q'}>\varepsilon^{(n-1)}_{q'}$. Using
similar arguments as in Case~A.4.a where $v_n$ is not an endpoint
in $\mathcal{P_{\mathcal{T}}}(G[S])$, we show that the algorithm
computes a minimum 1PC $\mathcal{P_{\mathcal{T}}}(G)$ of the graph
$G$ having $\varepsilon^{(n)}_i$ endpoints $v_\kappa$ belonging to
different paths with index $\kappa \in (i,n]$, $1 \leq i \leq n$,
such that there is no other minimum 1PC
$\mathcal{P'_{\mathcal{T}}}(G)$ having $\varepsilon'^{(n)}_i$
endpoints $v_{\kappa'}$ belonging to different paths with index
$\kappa' \in (i,n]$ such that
$\varepsilon'^{(n)}_i>\varepsilon^{(n)}_i$, $1 \leq i < \varrho$.
Furthermore, if $\varepsilon'^{(n)}_i = \varepsilon^{(n)}_i +1$,
$\varrho \leq i < \varrho'$ and $\varepsilon'^{(n)}_i =
\varepsilon^{(n)}_i$, $\varrho' \leq i \leq n$ then there exists a
vertex $v_z$ such that $\varepsilon^{(n)}_z >
\varepsilon'^{(n)}_z$, $1 \leq z < \varrho$ and there exists no
vertex $v_{z'}$, $1 \leq z' <\varrho$, such that
$\varepsilon'^{(n)}_{z'}>\varepsilon^{(n)}_{z'}$.

\y {\bf \textit{Case~A.6}:} When the algorithm processes vertex
$v_n$, it creates a new path having vertex $v_n$ as an endpoint,
that is, $\lambda_\mathcal{T}(G)=\lambda_\mathcal{T}(G[S])+1$.
This implies that $\forall i \geq d$ we have
$\varepsilon^{(n-1)}_i \leq 1$. Suppose that there exists a 1PC
$\mathcal{P'_{\mathcal{T}}}(G)$ of size
$\lambda_\mathcal{T}(G[S])-1$, that is, vertex $v_n$ is an
internal vertex of a path $P$ in $\mathcal{P'_{\mathcal{T}}}(G)$.
Then, removing vertex $v_n$ from $\mathcal{P'_{\mathcal{T}}}(G)$
results to a 1PC of $G[S]$ of size $\lambda_\mathcal{T}(G[S])$,
and, thus, minimum, such that there exists an index $i$ for which
$\varepsilon'^{(n-1)}_i=2$, a contradiction.

\y Suppose now that there exists a 1PC
$\mathcal{P'_{\mathcal{T}}}(G)$ of size
$\lambda_\mathcal{T}(G[S])$. Let $v_n$ be an endpoint of a path
$P$ in $\mathcal{P'_{\mathcal{T}}}(G)$. We distinguish the
following cases:

(i) $P=(v_n, v_r, \ldots, v_s)$. Removing vertex $v_n$ from
$\mathcal{P'_{\mathcal{T}}}(G)$ results to a 1PC of $G[S]$ of size
$\lambda_\mathcal{T}(G[S])$, and, thus, minimum, such that there
exists an index $i$ for which $\varepsilon'^{(n-1)}_i=1$, $i \geq
d$. Then, $d=d'$, which is equal to the index of the terminal
vertex, and $P$ is the terminal path such that its left endpoint
in $\mathcal{P'_{\mathcal{T}}}(G[S])$, that is, vertex $v_r$, has
index greater than the index of the left endpoint of the terminal
path in $\mathcal{P_{\mathcal{T}}}(G[S])$. Note that, $v_r$ cannot
be an endpoint in $\mathcal{P'_{\mathcal{T}}}(G[S])$ since
$\varepsilon'^{(n-1)}_i \leq \varepsilon^{(n-1)}_i$, $d \leq i
\leq n-1$, and, thus, a 1PC of $G[S]$ having $v_r$ as an endpoint
cannot be minimum. However, removing $v_n$ from
$\mathcal{P'_{\mathcal{T}}}(G)$ results to a 1PC of $G[S]$ of size
$\lambda_\mathcal{T}(G[S])$ having $v_r$ as an endpoint, a
contradiction.

(ii) $P=(v_t, v_n)$, where $v_t$ is the terminal vertex. Removing
$v_n$ and $v_t$ from $P$ results to a 1PC of $G[S]$ of size
$\lambda_\mathcal{T}(G[S])-1$, a contradiction. Indeed, since the
algorithm does not use $v_n$ to bridge two paths, removing $v_t$
from $G[S]$ results to $\lambda_\mathcal{T}(G[S])$ paths.

\y Now let $v_n$ be an internal vertex of a path $P=(v_r, \ldots,
v_i, v_n, v_j, \ldots, v_s)$ in $\mathcal{P'_{\mathcal{T}}}(G)$.
Suppose that $N(v_n)>0$ (the case where $N(v_n)=0$ is trivial) and
$v_t \notin N(v_n)$, where $v_t$ is the terminal vertex. Since the
algorithm constructs $\lambda_\mathcal{T}(G[S])+1$ paths, at least
$|N(v_n)|-1$ neighbors of $v_n$ have bridged paths reducing the
size of the 1PC and at most one of them was inserted; otherwise
there would exist at least two successive neighbors into a path or
at least one of them would be an endpoint. Suppose that $v_i$ and
$v_j$ have both bridged paths. Then, applying the algorithm to
$G-\{v_i,v_j,v_n\}$ would result to a minimum 1PC of
$G-\{v_i,v_j,v_n\}$ of size $\lambda_\mathcal{T}(G[S])+2$.
However, if we remove vertices $v_i, v_j$ and $v_n$ from
$\mathcal{P'_{\mathcal{T}}}(G)$ we obtain a 1PC of
$G-\{v_i,v_j,v_n\}$ of size $\lambda_\mathcal{T}(G[S])+1$, a
contradiction. Suppose now that $v_i$ was inserted and $v_j$ has
bridged paths. We distinguish the following cases:

(i) $j<i$. Clearly, applying the algorithm to $G'=G-\{v_i,v_n\}$
results to a minimum 1PC of $G'$ of size
$\lambda_\mathcal{T}(G[S])$. Furthermore, applying the algorithm
to $G'-\{v_j\}$ results to a minimum 1PC $\mathcal{P''}_1(G)$ of
$G'-\{v_j\}$ of size $\lambda_\mathcal{T}(G[S])+1$ such that no
free neighbor of $v_i$ is an endpoint and if $v_i$ sees the
terminal vertex $v_t$, it is not a trivial path in
$\mathcal{P''}_1(G)$. Indeed, any neighbor $v_a$ of $v_i$ such
that $i<a<n$ cannot be an endpoint in $\mathcal{P''}_1(G)$ since
every vertex $v_a$ such that $i<a<n$ is also a neighbor of $v_n$.
Note that, $t<j$. Furthermore, since $v_i$ is inserted, when
vertex $v_{j+1}$ was processed, no neighbor of $v_i$ was an
endpoint and if $v_iv_t \in E(G)$ vertex $v_t$ does not belong to
a trivial path. Indeed, let $v_k \in N(v_i)$ be an endpoint when
the algorithm processes vertex $v_{j+1}$ or $v_t$ belongs to a
trivial path. This implies that, when we apply the algorithm to
$G[S]$, we have one neighbor of $v_n$, say, $v_\ell$, bridging
through vertex $v_k$ or vertex $v_t$; then $v_i$ would be inserted
through the edge $v_kv_\ell$ or $v_tv_\ell$, which is a
contradiction since this results to two neighbors of $v_n$ being
successive. Additionally, no neighbor of $v_i$ becomes an endpoint
and vertex $v_t$ does not belong to a trivial path until vertex
$v_{n-1}$ is processed, since all vertices with index greater than
$j+1$ are neighbors of $v_n$, and, thus, they are used to bridge
paths reducing the size of the 1PC. Note that, according to the
algorithm, vertex $v_t$ cannot belong to a trivial path until
vertex $v_{n-1}$ is processed, since no bridge operation results
to $v_t$ belonging to a trivial path. Consequently, applying the
algorithm to $G'-\{v_j\}$ results to a minimum 1PC
$\mathcal{P''}_1(G'-\{v_j\})$ of size
$\lambda_\mathcal{T}(G[S])+1$ such that no free neighbor of $v_i$
is an endpoint and if $v_i$ sees the terminal vertex $v_t$, it is
not a trivial path in $\mathcal{P''}_1(G'-\{v_j\})$.

(ii) $i<j$. Similarly to case (i), applying the algorithm to
$G'-\{v_j\}$ results to a minimum 1PC
$\mathcal{P''}_1(G'-\{v_j\})$ of size
$\lambda_\mathcal{T}(G[S])+1$ such that no free neighbor of $v_i$
is an endpoint and if $v_i$ sees the terminal vertex $v_t$, it is
not a trivial path in $\mathcal{P''}_1(G'-\{v_j\})$. Indeed, when
vertex $v_{i+1}$ is processed, no neighbor of $v_i$ is an endpoint
and if $v_i$ sees the terminal vertex $v_t$, it is not a trivial
path. Furthermore, since no neighbor of $v_n$ can be an endpoint,
no vertex with index greater than $i+1$ is an endpoint.
Additionally, no neighbor of $v_i$ becomes an endpoint and if
$v_tv_i \in E(G)$, vertex $v_t$ does not belong to a trivial path
until vertex $v_{n-1}$ is processed, since all vertices with index
greater than $i+1$ are neighbors of $v_n$, and, thus, they are
used to bridge paths reducing the size of the 1PC. Note that,
according to the algorithm, vertex $v_t$ cannot belong to a
trivial path until vertex $v_{n-1}$ is processed. Consequently,
applying the algorithm to $G'-\{v_j\}$ results to a minimum 1PC
$\mathcal{P''}_1(G'-\{v_j\})$ of size
$\lambda_\mathcal{T}(G[S])+1$ such that no free neighbor of $v_i$
is an endpoint and if $v_i$ sees the terminal vertex $v_t$, it is
not a trivial path in $\mathcal{P''}_1(G'-\{v_j\})$.

\y Since $v_n$ is an internal vertex of a path $P=(v_r, \ldots,
v_i, v_n, v_j, \ldots, v_s)$ in $\mathcal{P'_{\mathcal{T}}}(G)$
which has size $\lambda_\mathcal{T}(G[S])$, if we remove vertices
$v_i, v_j$ and $v_n$ from $P$ we obtain a 1PC of $G'-\{v_j\}$ of
size $\lambda_\mathcal{T}(G[S])+1$ such that a free neighbor of
$v_i$ is an endpoint, a contradiction; the same holds when
$P=(v_r, \ldots, v_i, v_n,$ $v_j, v_t)$. If $P=(v_t, v_i, v_n,
v_j, \ldots, v_s)$ then $v_t$ belongs to a trivial path in
$\mathcal{P''}_1(G)$, a contradiction. If $P=(v_i, v_n, v_j,
\ldots, v_s)$ or $P=(v_r, \ldots, v_i, v_n, v_j)$, then removing
vertices $v_i, v_j$ and $v_n$ from $P$ results to a 1PC of
$G'-\{v_j\}$ of size $\lambda_\mathcal{T}(G[S])$, a contradiction.

\y Now let $v_t \in N(v_n)$, where $v_t$ is the terminal vertex.
The case where $v_n$ is an internal vertex of a path $P=(v_r,
\ldots, v_i, v_n, v_j, \ldots, v_s)$ in
$\mathcal{P'_{\mathcal{T}}}(G)$ which has size
$\lambda_\mathcal{T}(G[S])$ leads to a contradiction similarly to
the case where $v_t \notin N(v_n)$. Suppose that $v_n$ is an
internal vertex of a path $P=(v_t, v_n, v_j, \ldots, v_s)$ in
$\mathcal{P'_{\mathcal{T}}}(G)$ which has size
$\lambda_\mathcal{T}(G[S])$. According to the algorithm, no
neighbor of $v_n$ is inserted until vertex $v_t$ is processed.
Also, it is easy to see that, no neighbor of $v_n$ with index
greater than $t$ is inserted, either. Indeed, let $v_a$, $t<a<n$,
be a neighbor of $v_n$ which is inserted into the terminal path.
Since the algorithm results to $\lambda_\mathcal{T}(G[S])+1$
paths, $v_a$ is inserted through an edge $v_kv_\ell$ such that
$v_k,v_\ell \notin N(v_n)$ and $v_t$ is connected to a vertex
$v_q$ such that $v_q \notin N(v_n)$. This implies that the
ordering of the vertices $v_t, v_k, v_\ell$ and $v_q$ of the
terminal path is as follows: $v_q<v_k<v_\ell<v_t$ or
$v_q<v_\ell<v_k<v_t$; Without loss of generality suppose that
$v_q<v_k<v_\ell<v_t$. Let $v_b$ be the other endpoint of the
terminal path. Clearly $v_b<v_k$. Also, without loss of
generality, suppose that $v_b<v_q$. Consequently, when vertex
$v_t$ is processed, the algorithm has constructed a path having
two successive vertices, $v_k$ and $v_\ell$, which have indexes
greater than those of the endpoints of the path, that is, $v_b$
and $v_q$. This is a contradiction, since it implies that there
exists at least one vertex with index greater than $q$ which sees
$v_q$; in this case the algorithm could not result to a path
having $v_q$ as an endpoint. Consequently, we have shown that if
we apply the algorithm to $G[S]$, no neighbor of $v_n$ is inserted
into the terminal path. Furthermore, since there are no neighbors
of $v_n$ successive into a path, all neighbors of $v_n$ bridge
paths reducing the size of the 1PC. This implies that, if we apply
the algorithm to $G[S]-\{v_t\}$, we obtain a minimum 1PC of
$G[S]-\{v_t\}$ of size $\lambda_\mathcal{T}(G[S])$. Furthermore,
if we apply the algorithm to $G[S]-\{v_t,v_j\}$, we obtain a
minimum 1PC of $G[S]-\{v_t,v_j\}$ of size
$\lambda_\mathcal{T}(G[S])+1$. However, removing $v_t, v_n$ and
$v_j$ from $P=(v_t, v_n, v_j, \ldots, v_s)$ which is a path in
$\mathcal{P'_{\mathcal{T}}}(G)$, we obtain a 1PC of
$G[S]-\{v_t,v_j\}$ of size at most $\lambda_\mathcal{T}(G[S])$, a
contradiction; thus, $v_n$ cannot be an internal vertex of a path
$P=(v_t, v_n, v_j, \ldots, v_s)$ in
$\mathcal{P'_{\mathcal{T}}}(G)$ which has size
$\lambda_\mathcal{T}(G[S])$.

\y We have shown that there does not exist a 1PC
$\mathcal{P'_{\mathcal{T}}}(G)$ of size
$\lambda_\mathcal{T}(G[S])$, and, thus,
$\mathcal{P_{\mathcal{T}}}(G)$ is a minimum 1PC of $G$, that is,
$\lambda'_\mathcal{T}(G)=\lambda_\mathcal{T}(G)=\lambda_\mathcal{T}(G[S])+1$.

\y Using similar arguments as in Case~A.4.a where $v_n$ is an
endpoint in $\mathcal{P_{\mathcal{T}}}(G[S])$, we show that the
algorithm computes a minimum 1PC $\mathcal{P_{\mathcal{T}}}(G)$ of
every interval graph $G$ with $n$ vertices having
$\varepsilon^{(n)}_i$ endpoints $v_\kappa$ belonging to different
paths with index $\kappa \in (i,n]$, $1 \leq i \leq n$, such that
there is no other minimum 1PC $\mathcal{P'_{\mathcal{T}}}(G)$
having $\varepsilon'^{(n)}_i$ endpoints $v_{\kappa'}$ belonging to
different paths with index $\kappa' \in (i,n]$ such that
$\varepsilon'^{(n)}_i>\varepsilon^{(n)}_i$, $1 \leq i \leq n$.

\yy {\bf Case~B:} vertex $v_n$ is the terminal vertex. Clearly,
the size $\lambda'_\mathcal{T}(G)$ of a minimum 1PC of $G$ is
equal to $\lambda_\mathcal{T}(G[S])$ or
$\lambda_\mathcal{T}(G[S])+1$. We distinguish the following cases:

\y {\bf \textit{Case~B.1}:} When the algorithm processes vertex
$v_n$, it connects $v_n$ to a path, that is,
$\lambda_\mathcal{T}(G)=\lambda_\mathcal{T}(G[S])$. Since $v_n$ is
the terminal vertex, the 1PC $\mathcal{P_{\mathcal{T}}}(G)$ is a
minimum 1PC of $G$, that is,
$\lambda'_\mathcal{T}(G)=\lambda_\mathcal{T}(G)=\lambda_\mathcal{T}(G[S])$.

\y We show that the algorithm computes a minimum 1PC
$\mathcal{P_{\mathcal{T}}}(G)$ of the graph $G$ having
$\varepsilon^{(n)}_i$ endpoints $v_\kappa$ belonging to different
paths with index $\kappa \in (i,n]$, $1 \leq i \leq n$, such that
there is no other minimum 1PC $\mathcal{P'_{\mathcal{T}}}(G)$
having $\varepsilon'^{(n)}_i$ endpoints $v_{\kappa'}$ belonging to
different paths with index $\kappa' \in (i,n]$ such that
$\varepsilon'^{(n)}_i>\varepsilon^{(n)}_i$, $1 \leq i \leq n$.

\y Suppose that $v_n \in P=(v_n, v_a, \ldots) \in
\mathcal{P_{\mathcal{T}}}(G)$. Then, there cannot exist a minimum
1PC $\mathcal{P'_{\mathcal{T}}}(G)$ such that $1=
\varepsilon^{(n)}_{\varrho-1} < \varepsilon'^{(n)}_{\varrho-1}$.
Assume that there exists a minimum 1PC
$\mathcal{P'_{\mathcal{T}}}(G)$ having an index, say, $k-1$, for
which we have $\varepsilon'^{(n)}_{k-1}$ endpoints $v_{\kappa'}$
belonging to different paths with index $\kappa' \in (k-1,n]$,
where $\varepsilon'^{(n)}_{k-1}>\varepsilon^{(n)}_{k-1}$.
Similarly, to Case~A.1, there exists a vertex $v_j$, $1 \leq j <
k-1$, such that $\varepsilon^{(n)}_j = \varepsilon'^{(n)}_j$.

\y Suppose that $v_n \in P'=(v_n, v_{a'}, \ldots) \in
\mathcal{P'_{\mathcal{T}}}(G)$. If $v_{a'}<v_{j+1}$, then
$v_{j+1}v_n \in E(G)$, and, thus, $v_a<v_{j+1}$. Then, if we
remove $v_n$ from $\mathcal{P_{\mathcal{T}}}(G)$ and
$\mathcal{P'_{\mathcal{T}}}(G)$, we obtain
$\varepsilon^{(n-1)}_{j+1}=\varepsilon^{(n)}_{j+1}-1=\varepsilon'^{(n)}_{j+1}-2$
and $\varepsilon'^{(n-1)}_{j+1}=\varepsilon'^{(n)}_{j+1}-1$; thus,
$\varepsilon^{(n-1)}_{j+1}=\varepsilon'^{(n-1)}_{j+1}-1$, a
contradiction. On the other hand, if $v_{a'}>v_{j+1}$, then, if we
remove $v_n$ from $\mathcal{P_{\mathcal{T}}}(G)$ and
$\mathcal{P'_{\mathcal{T}}}(G)$, we obtain
$\varepsilon^{(n-1)}_{j+1}=\varepsilon^{(n)}_{j+1}-1=\varepsilon'^{(n)}_{j+1}-2$
or
$\varepsilon^{(n-1)}_{j+1}=\varepsilon^{(n)}_{j+1}=\varepsilon'^{(n)}_{j+1}-1$.
Also, $\varepsilon'^{(n-1)}_{j+1}=\varepsilon'^{(n)}_{j+1}$; thus,
$\varepsilon^{(n-1)}_{j+1}=\varepsilon'^{(n-1)}_{j+1}-1$, a
contradiction.

\y {\bf \textit{Case~B.2}:} When the algorithm processes vertex
$v_n$, it constructs a new trivial path, that is,
$\lambda_\mathcal{T}(G)=\lambda_\mathcal{T}(G[S])+1$. Suppose that
there exists a 1PC $\mathcal{P'_{\mathcal{T}}}(G)$ of size
$\lambda_\mathcal{T}(G[S])$. Clearly, vertex $v_n$ cannot belong
to a trivial path in $\mathcal{P'_{\mathcal{T}}}(G)$, since
removing it results to a 1PC of $G[S]$ of size
$\lambda_\mathcal{T}(G[S])-1$, a contradiction. Thus, let $P=(v_n,
v_r, \ldots) \in \mathcal{P'_{\mathcal{T}}}(G)$ be the path
containing $v_n$. Removing vertex $v_n$ from
$\mathcal{P'_{\mathcal{T}}}(G)$ results to a 1PC of $G[S]$ of size
$\lambda_\mathcal{T}(G[S])$, and, thus, minimum, having a neighbor
of $v_n$, that is, vertex $v_r$, as an endpoint of a path. Since
$G[S]$ does not contain the terminal vertex, according to the
induction hypothesis, this is a contradiction. Consequently, the
1PC $\mathcal{P_{\mathcal{T}}}(G)$ is a minimum 1PC of $G$, that
is,
$\lambda'_\mathcal{T}(G)=\lambda_\mathcal{T}(G)=\lambda_\mathcal{T}(G[S])+1$.

\y We show that the algorithm computes a minimum 1PC
$\mathcal{P_{\mathcal{T}}}(G)$ of the graph $G$ having
$\varepsilon^{(n)}_i$ endpoints $v_\kappa$ belonging to different
paths with index $\kappa \in (i,n]$, $1 \leq i \leq n$, such that
there is no other minimum 1PC $\mathcal{P'_{\mathcal{T}}}(G)$
having $\varepsilon'^{(n)}_i$ endpoints $v_{\kappa'}$ belonging to
different paths with index $\kappa' \in (i,n]$ such that
$\varepsilon'^{(n)}_i>\varepsilon^{(n)}_i$, $1 \leq i \leq n$.

\y Since $P=(v_n)$ there cannot exist a minimum 1PC
$\mathcal{P'_{\mathcal{T}}}(G)$ such that $1=
\varepsilon^{(n)}_{\varrho-1} < \varepsilon'^{(n)}_{\varrho-1}$.
Assume that there exists a minimum 1PC
$\mathcal{P'_{\mathcal{T}}}(G)$ having an index, say, $k-1$, for
which we have $\varepsilon'^{(n)}_{k-1}$ endpoints $v_{\kappa'}$
belonging to different paths with index $\kappa' \in (k-1,n]$,
where $\varepsilon'^{(n)}_{k-1}>\varepsilon^{(n)}_{k-1}$. Suppose
that $\varepsilon^{(n)}_{k-1}=x$ and $\varepsilon'^{(n)}_{k-1} =
x+1$. Similarly, to Case~A.1, there exists a vertex $v_j$, $1 \leq
j < k-1$, such that $\varepsilon^{(n)}_j = \varepsilon'^{(n)}_j$.

\y Suppose that $v_n \in P'=(v_n, v_{a'}, \ldots) \in
\mathcal{P'_{\mathcal{T}}}(G)$; the case where $P'=(v_n)$ is
trivial. If $v_{a'}<v_{j+1}$, then $v_{j+1}v_n \in E(G)$, and,
thus, vertex $v_n$ would be connected, a contradiction. If
$v_{a'}>v_{j+1}$, then, if we remove $v_n$ from
$\mathcal{P_{\mathcal{T}}}(G)$ and
$\mathcal{P'_{\mathcal{T}}}(G)$, we obtain
$\varepsilon^{(n-1)}_{j+1}=\varepsilon^{(n)}_{j+1}-1=\varepsilon'^{(n)}_{j+1}-2$
and $\varepsilon'^{(n-1)}_{j+1}=\varepsilon'^{(n)}_{j+1}$, a
contradiction. \s \qed

\yy Let $G=(V,E)$ be an interval graph on $n$ vertices and $m$
edges and let $\mathcal{T}$ be a terminal set containing a vertex
$v \in V(G)$. Then, Algorithm Minimum\_1PC computes a minimum 1PC
$\mathcal{P}_\mathcal{T}(G)$ of $G$ in $O(n^2)$ time and requires
linear space. Recall that the ordering $\pi$ of the vertices is
constructed in linear time \cite{RamRan}. Hence, we can state the
following result.

\bigskip
\par\noindent
{\bf Theorem~4.2.} {\it Let $G$ be an interval graph on
$n$~vertices and let $\mathcal{T}$ be a subset of $V(G)$
containing a single vertex. A minimum 1-fixed-endpoint path cover
of $G$ with respect to $\mathcal{T}$ can be computed in $O(n^2)$
time.}

\vskip 0.3in 
\section{Related Results on Convex and Biconvex Graphs}
Based on the results for the 1PC problem on interval graphs, and
also on the reduction described by M\"{u}ller in \cite{Muller}, we
study the HP and 1HP problems on convex and biconvex graphs. A
bipartite graph $G=(X,Y;E)$ is called {\it $X$-convex} (or simply
convex) if there exists an ordering $<$ so that for all $y \in Y$
the set N(y) is $<$-consecutive \cite{Muller}; $G$ is {\it
biconvex} if it is convex on both $X$ and $Y$.

\y In this section, we solve the HP and 1HP problems on a biconvex
graph $G=(X,Y;E)$. Moreover, we show that the HP problem on a
$X$-convex graph $G(X,Y;E)$ on $n$ vertices can be solved in
$O(n^3)$ time if $|X|=|Y|$ or $|X|-|Y|=1$ and a 1HP starting at
vertex $u$, if there exists, can be computed in $O(n^2)$ time if
($|X|=|Y|$ and $u \in Y$) or $|X|-|Y|=1$.

\y  We next describe an algorithm for the HP problem on a biconvex
graph $G=(X,Y;E)$. Note that the operation Algorithm\_HP
corresponds to the algorithm for computing a minimum path cover of
an interval graph described in \cite{AR90}.

\bigskip \noindent{\it Algorithm HP\_Biconvex} \y \noindent{\it
Input:} a biconvex graph $G=(X,Y;E)$ on $n$ vertices; \y
\noindent{\it Output:} a Hamiltonian path of $G$, if one exists;
\y

\begin{enumerate}
  \item {\tt if} $||X|-|Y||>1$ {\tt then} {\tt return}($G$ does not have a Hamiltonian path);
  \item {\tt if} $|X|=|Y|$ {\tt then} \\
         \phantom{if} {\tt construct} the interval graph $G'$: $V(G')=X \cup Y$, $E(G')=E \cup E_Y$,
         where $E_Y$ is as follows: \\
         \phantom{if} \phantom{if} \phantom{if} $\{y_1y_2 \in E_Y\}$ iff $y_1,y_2 \in Y$ and $N(y_1) \cap N(y_2) \neq \emptyset$;\\
         \phantom{if} {\tt if} $\exists y_j \in Y:|N(y_j)|=1$ {\tt then} \\
         \phantom{if} \phantom{if} \phantom{if} $\mathcal{P}_\mathcal{T}(G)=${\tt Minimum\_1PC($G'$,$y_j$)};\\
         \phantom{if} \phantom{if} \phantom{if} {\tt if} $\lambda_\mathcal{T}(G)=1$
         {\tt then} {\tt return($\mathcal{P}_\mathcal{T}(G)$);}\\
         \phantom{if} \phantom{if} \phantom{if} {\tt else} {\tt return}($G$ does not have a Hamiltonian path);\\
         \phantom{if} {\tt else} \\
         \phantom{if} \phantom{if} \phantom{if} {\tt for} $i=1$ {\tt to} $|Y|$ {\tt do} \\
         \phantom{if} \phantom{if} \phantom{if} \phantom{for} $\mathcal{P}_\mathcal{T}(G)=$ {\tt Minimum\_1PC($G',y_i$)};\\
         \phantom{if} \phantom{if} \phantom{if} \phantom{for} {\tt if} $\lambda_\mathcal{T}(G)=1$ {\tt then} {\tt return($\mathcal{P}_\mathcal{T}(G)$);}\\
         \phantom{if} \phantom{if} \phantom{if} {\tt end-for;}\\
         \phantom{if} \phantom{if} \phantom{if} {\tt return}($G$ does not have a Hamiltonian path);
  \item {\tt if} $|X|-|Y|=1$ {\tt then} \\
         \phantom{if} $\mathcal{P}_\mathcal{T}(G)=${\tt Algorithm\_HP($G'$)};\\
         \phantom{if} {\tt if} $\lambda_\mathcal{T}(G)=1$ {\tt then} {\tt return($\mathcal{P}_\mathcal{T}(G)$);}\\
         \phantom{if} {\tt else} {\tt return}($G$ does not have a Hamiltonian path);
  \item {\tt if} $|Y|-|X|=1$ {\tt then} \\
         \phantom{if} {\tt construct} the interval graph $G'$: $V(G')=X \cup Y$, $E(G')=E \cup E_X$,
         where $E_X$ is as follows: \\
         \phantom{if} \phantom{if} $\{x_1x_2 \in E_X\}$ iff $x_1,x_2 \in X$ and $N(x_1) \cap N(x_2) \neq \emptyset$;\\
         \phantom{if} $\mathcal{P}_\mathcal{T}(G)=${\tt Algorithm\_HP($G'$)};\\
         \phantom{if} {\tt if} $\lambda_\mathcal{T}(G)=1$ {\tt then} {\tt return($\mathcal{P}_\mathcal{T}(G)$);}\\
         \phantom{if} {\tt else} {\tt return}($G$ does not have a Hamiltonian path);
\end{enumerate}

\y \noindent {\bf Observation~5.1.} \s Uehara and Uno in
\cite{UeharaUno} claim that the HP problem on a biconvex graph
$G(X,Y;E)$ on $n$ vertices can be solved in $O(n^2)$ time even if
$|X|=|Y|$. Specifically, they claim that $G$ has an HP if and only
if the interval graph $G'$ has an HP, where $G'$ is an interval
graph such that $V(G')=X \cup Y$ and $E(G')=E \cup E_Y$, where
$E_Y$ is as follows: $\{y_1y_2 \in E_Y\}$ iff $y_1,y_2 \in Y$ and
$N(y_1) \cap N(y_2) \neq \emptyset$. However, this is not true,
since there exists a counterexample, which is presented at
Figure~\ref{biconvex}. Indeed, the biconvex graph $G$ of
Figure~\ref{biconvex} does not have an HP while for $G'$ we have
$P=(x_1, y_2, x_2, y_1, y_4, x_3, y_3, x_4)$. Suppose that we
construct an interval graph $G'$ such that $V(G')=X \cup Y$ and
$E(G')=E \cup E_X$, where $E_X$ is as follows: $\{x_1x_2 \in
E_X\}$ iff $x_1,x_2 \in X$ and $N(x_1) \cap N(x_2) \neq
\emptyset$. Then, $G'$ has an HP, that is, $P=(y_4, x_3, y_3, x_4,
x_1, y_2, x_2, y_1)$. Thus, there exists no algorithm with time
complexity $O(n^2)$ and we can state the following result.

\begin{figure}[t]
\yy \hrule \y\y\y
  \centering
  \includegraphics[scale=1]{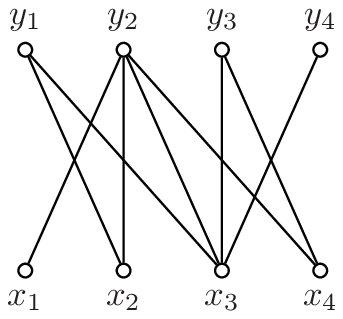}
  \centering
  \caption{\small{A biconvex graph $G$.}}
  \label{biconvex}
 \yy \hrule \y\y
\end{figure}

\bigskip
\par\noindent
{\bf Theorem~5.1.} {\it The Hamiltonian path problem on a biconvex
graph $G$ on $n$ vertices can be solved in $O(n^3)$ time.}

\bigskip Similarly, we show that the Hamiltonian path problem on a
$X$-convex graph $G(X,Y;E)$ on $n$ vertices can be solved in
$O(n^3)$ time when $|X|=|Y|$ or $|X|-|Y|=1$. It is easy to see
that if $|Y|-|X|=1$ then the $X$-convex graph $G(X,Y;E)$ has a
Hamiltonian path if and only if the interval graph $G'$ has a 2HP
between any two vertices of $Y$. Thus, we can state the following
result.

\bigskip
\par\noindent
{\bf Corollary~5.1.} {\it The Hamiltonian path problem on a
$X$-convex graph $G(X,Y;E)$ on $n$ vertices can be solved in
$O(n^3)$ time if $|X|=|Y|$ or $|X|-|Y|=1$.}

\bigskip We next describe an algorithm for the 1HP problem on a
biconvex graph $G=(X,Y;E)$.

\bigskip \noindent{\it Algorithm 1HP\_Biconvex} \y \noindent{\it
Input:} a biconvex graph $G=(X,Y;E)$ on $n$ vertices and a vertex
$y_t \in Y)$; \y \noindent{\it Output:} a a Hamiltonian path of
$G$ starting at vertex $y_t$, if one exists; \y

\begin{enumerate}
  \item {\tt if} $||X|-|Y||>1$ {\tt then} {\tt return}($G$ does not have a 1HP);
  \item {\tt if} $|X|=|Y|$ {\tt then} \\
         \phantom{if} {\tt construct} the interval graph $G'$: $V(G')=X \cup Y$, $E(G')=E \cup E_Y$, where $E_Y$ is as follows: \\
         \phantom{if} \phantom{if} \phantom{if} $\{y_1y_2 \in E_Y\}$ iff $y_1,y_2 \in Y$ and $N(y_1) \cap N(y_2) \neq \emptyset$;\\
         \phantom{if} $\mathcal{P}_\mathcal{T}(G)=${\tt Minimum\_1PC($G', y_t$)};\\
         \phantom{if} {\tt if} $\lambda_\mathcal{T}(G)=1$ {\tt then} {\tt return($\mathcal{P}_\mathcal{T}(G)$);}\\
         \phantom{if} {\tt else} {\tt return}($G$ does not have a 1HP);
  \item {\tt if} $|X|-|Y|=1$ {\tt then} {\tt return}($G$ does not have a 1HP);
  \item {\tt if} $|Y|-|X|=1$ {\tt then} \\
         \phantom{if} {\tt construct} the interval graph $G'$: $V(G')=X \cup Y$, $E(G')=E \cup E_X$, where $E_X$ is as follows: \\
         \phantom{if} \phantom{if} $\{x_1x_2 \in E_X\}$ iff $x_1,x_2 \in X$ and $N(x_1) \cap N(x_2) \neq \emptyset$;\\
         \phantom{if} $\mathcal{P}_\mathcal{T}(G)=${\tt Minimum\_1PC($G', y_t$)};\\
         \phantom{if} {\tt if} $\lambda_\mathcal{T}(G)=1$ {\tt then} {\tt return($\mathcal{P}_\mathcal{T}(G)$);}\\
         \phantom{if} {\tt else} {\tt return}($G$ does not have a 1HP);
  \end{enumerate}

\y Since Algorithm Minimum\_1PC requires $O(n^2)$ time to compute
a 1HP of an interval graph on $n$ vertices and the graph $G'$ can
be constructed in $O(|X \cup Y|^2)$ time \cite{Muller}, Algorithm
1HP\_Biconvex returns a 1HP, if there exists, of a biconvex graph
on $n$ vertices in $O(n^2)$ time. Hence, we can state the
following result.

\bigskip
\par\noindent
{\bf Theorem~5.2.} {\it Let $G$ be a biconvex graph on
$n$~vertices and let $u$ be a vertex of $V(G)$. The 1HP problem on
$G$ can be solved in $O(n^2)$ time.}

\bigskip Let $G(X,Y;E)$ be a $X$-convex graph on
$n$~vertices and let $u$ be a vertex of $V(G)$. Similarly, we show
that the 1HP problem on $G$ can be solved in $O(n^2)$ time when
($|X|=|Y|$ and $u \in Y$) or $|X|-|Y|=1$. Clearly, if $|Y|-|X|=1$
and $u \in X$ then $G$ does not have a 1HP. It is easy to see that
if ($|Y|-|X|=1$ and $u \in Y$) or ($|X|=|Y|$ and $u \in X$) then
the $X$-convex graph $G(X,Y;E)$ has a Hamiltonian path if and only
if the interval graph $G'$ has a 2HP between $u$ and a vertex of
$Y$. Thus, we can state the following result.

\bigskip
\par\noindent
{\bf Corollary~5.2.} {\it Let $G(X,Y;E)$ be a $X$-convex graph on
$n$~vertices and let $u$ be a vertex of $V(G)$. The 1HP problem on
$G$ can be solved in $O(n^2)$ time if ($|X|=|Y|$ and $u \in Y$) or
$|X|-|Y|=1$. If $|Y|-|X|=1$ and $u \in X$ then $G$ does not have a
1HP.}

\vskip 0.3in 
\section{Concluding Remarks}

This paper presents an $O(n^2)$ time algorithm for the 1PC problem
on interval graphs. Given an interval graph $G$ and a vertex $v$
of $G$, our algorithm constructs a minimum path cover of $G$ such
that $v$ is an endpoint. Thus, if the graph $G$ is Hamiltonian,
our algorithm constructs a 1HP. It would be interesting to see if
the problem can be solved in linear time. Furthermore, an
interesting open question is whether the $k$-fixed-endpoint path
cover problem (kPC) can be polynomially solved on interval graphs.
Given a graph $G$ and a subset $\mathcal{T}$ of $k$ vertices of
$V(G)$, a $k$-fixed-endpoint path cover of $G$ with respect to
$\mathcal{T}$ is a set of vertex-disjoint paths $\mathcal{P}$ that
covers the vertices of $G$ such that the $k$ vertices of
$\mathcal{T}$ are all endpoints of the paths in $\mathcal{P}$. The
kPC problem is to find a $k$-fixed-endpoint path cover of $G$ of
minimum cardinality. Note that, the kPC problem generalizes the
2HP problem; the complexity status of the 2HP problem on interval
graphs remains an open question. }

\end{document}